\documentclass[]{aa}
\usepackage{natbib}
\bibpunct{(}{)}{;}{a}{}{,}
\usepackage[pdftex]{graphicx,hyperref}
 \usepackage{subfigure}
\usepackage{txfonts}

\title{The application of compressive sampling to radio astronomy}
\subtitle{\textrm{I}: Deconvolution}

\author{F. Li \inst{1}, T. J. Cornwell \inst{1} , F. de Hoog \inst{1} }
\offprints{F. Li (feng.li@csiro.au)}
\institute{Commonwealth Scientific and Industrial Research Organization (CSIRO), Australia}

\begin{document}

\abstract{Compressive sampling is a new paradigm for sampling, based on sparseness of signals or signal representations. It is much less restrictive than Nyquist-Shannon sampling theory and thus explains and systematises the widespread experience that methods such as the H\"ogbom CLEAN can violate the Nyquist-Shannon sampling requirements.  In this paper, a CS-based deconvolution method for extended sources is introduced. This method can reconstruct both point sources and extended sources (using the isotropic undecimated wavelet transform as a basis function for the reconstruction step). We compare this CS-based deconvolution method with two CLEAN-based deconvolution methods: the H\"ogbom CLEAN and the multiscale CLEAN. This new method shows the best performance in deconvolving extended sources for both uniform and natural weighting of the sampled visibilities. Both visual and numerical results of the comparison are provided. }

\date{Received date; accepted date}
\titlerunning{CS deconvolution}
\maketitle

\section{Introduction}

\label{sec:intro} 

Radio interferometers measure the spatial coherence function of the electric field \citep{Thompson:1986p2482}. From complete sampling of this function, an image may be constructed by applying an inverse Fourier transform. Nyquist-Shannon sampling theory dictates the sampling required. However, since the development of the CLEAN algorithm~\citep{Hogbom:1974p1206}, it has been known that given some restrictions on the sky brightness - specifically that it is composed of a limited number of point sources - the Nyquist-Shannon limits are much too conservative. 

Compressive sensing/sampling theory (CS)~\citep{Candes:2008p14, Candes:2006p23,Wakin:2008p1623} says that we can reconstruct a signal using far fewer measurements than required by the Nyquist-Shannon theory, provided that the signal is sparse or there is a sparse representation of the signal with a respective given basis function dictionary. Since CS was proposed, it has attracted very substantial interest and been applied in many research areas including: a single 
pixel camera~\citep{Wakin:2006p1437}, a high performance magnetic resonance imaging~\citep
{Lustig:2007p1719,Puy:2010p1807}, a modulated wideband converter~\citep
{Mishali:2009p2147}, image codec on Herschel satellite~\citep{Bobin:2009p2006}, and so on. 

CS can be applied to radio interferometry. \cite{Wiaux:2009p2267} compare the CS-based deconvolution methods with the H\"ogbom CLEAN method~\citep{Hogbom:1974p1206} on simulated uniform random sensing matrices with different coverage rates. They apply compressive sensing straightforwardly for deconvolution by assuming that the target signal is sparse. In their paper, they did not consider a sparse representation that is appropriate for extended sources in the sky. In a subsequent paper, Wiaux and other authors propose a new spread spectrum technique for radio interferometry~\citep{Wiaux:2009p1697} by using the non-negligible and constant component of the antenna separation in the pointing direction. Moreover, in their second paper, they assumed that any kind of astrophysical structure consist of Gaussian waveforms of equal size and similar standard deviation. Neither of these two papers provide a complete comparison between the CS-based deconvolution methods and the CLEAN-based deconvolution methods, such as the H\"ogbom CLEAN and the multiscale CLEAN~\citep{Cornwell:2008p191}. In this paper, we introduce a CS-based deconvolution method that deals with extended emission without assuming Gaussian shapes. This method adopts the isotropic undecimated wavelet transform (IUWT)~\citep{Starck:2007p1136} as a basis function. Our motivation for adopting the IUWT is based upon the work of Starck $et. al$, who found that the IUWT can decompose a wide range of astronomical images into a sparse isotropic representation \citep{Starck:2007p1136}.  We compare the performance of the resulting deconvolution method with H\"ogbom CLEAN, the multiscale CLEAN~\citep{Cornwell:2008p191}, and another CS-based method. Both visual and numerical results indicate that the new CS-based deconvolution method provides superior results regardless of the uniform or natural weighting adopted.

In Section~\ref{sec:csintro}, we review compressive sensing theory. The working mechanism of CS and its critical ingredients such as sparsity, incoherence, the restricted isometry property, and reconstruction are introduced. In Section~\ref{sec:radiointro}, we introduce the fundamentals of radio astronomy. In Section~\ref{sec:csdeconvolution}, the CS-based deconvolution methods are discussed. Our visual and numerical comparisons of deconvolution methods in Section~\ref{sec:results}.  The final conclusions are given in Section~\ref{sec:conclusion}.

\section{Compressive sensing}
\label{sec:csintro} 

As mentioned in the introduction, CS theory governs the sensing or sampling of a sparse signal. In this paper, we are concerned with the application of CS to images. We begin by considering image compression. In many image compression techniques, images are transformed into other domains where a limited number of parameters can represent the image. For example, the wavelet transform can provide a sparse representation of real-world images, because there are many small values on each scale for the smooth regions of the images, and there are a few large values for the edges. The fundamental technique in modern image compression ~\citep{Shapiro:1993p2541, Said1996} is to retain only the large-value wavelet coefficients. The wavelet transform was adopted for the JPEG2000 standard for image compression. In this procedure, images are captured by an imaging system, transformed into a wavelet domain, small or insignificant values discarded, and the transform reversed. While this works well, the approach is wasteful since too much information is recorded in the first place. CS provides a superior strategy in which the measurements themselves are limited or sparse. For examples, two CS-based layouts of CMOS sensors were proposed in~\citet{Robucci2008} and \citet{Bibet:2009p2787}. With these CMOS sensors, one can reduce the power drawn from commercial camera batteries by discarding the requirement for codec chips to be able to perform compression. Another application is to trade off the compression rate against number of measurements~\citep{Shapiro:1993p2541, Said1996}.
 
To describe CS concisely, we describe a general linear imaging system with
 \begin{equation}\label{equ_Isparse}
 \psi I=V,
\end{equation} where $I$ is the target signal in vector form of length $n$, the vector $V$ is  the observations with length $m$, and $\psi $ is a matrix of size $m$ by $n$. If there are fewer observations than unknowns in the 
target signal $I$, $i.e.$, there are more columns than 
rows in $\psi$ and $m\ll n$, we must try to recover the unknowns in the target signal with fewer observations. Although this equation may admit multiple solutions, if 
vector signal $I$ is sufficiently sparse with respect to a known dictionary of basis functions, we can fully reconstruct the target signal $I$  by using this extra information. That this is possible is known from the success of non-linear imaging algorithms such as the H\"ogbom CLEAN. The advance in CS is a rigorous theory constraining under what circumstances this is possible.  

Compressive sensing theory has a number of important ingredients: sparsity, incoherence, the restricted isometry property, and reconstruction. We introduce these in the next subsections.

\subsection{Sparsity}
\label{sec:cs_sparsity} 
A signal is sparse if there is is a given dictionary of basis functions in which the signal can be represented by most elements being zero. Examples of dictionaries of basis functions include the Fourier transform, the wavelet transform, and the gradient representation for a piecewise constant signal. With a sparse basis function, Eq. (\ref{equ_Isparse}) can be rewritten as
 \begin{equation}
\label{equ_basissparse}
 \psi\phi\alpha=V  \; \; \mathrm{or}\;  \;\Theta\alpha=V,
\end{equation} where $\Theta= \psi\phi$ of size $m\times n$ and $I=\phi\alpha$, $\alpha$ is a 
sparse representation when the transform matrix $\phi$ is adopted. In general, $\phi$ is $n\times l$ with $l$ greater or equal to $n$. This allows compressive sensing to have a wide application, because many signals often have a sparse representation in a certain
basis dictionary. This can, of course, also be applied to the circumstances where the signal itself can be sparse, in which case $\phi$ is the identity matrix.   

\subsection{Incoherence}
\label{sec:cs_incoherent} 
 
Nyquist-Shannon sampling theory covers the requirements when sampling the signal directly. In contrast,  CS theory covers sensing or 
sampling of the target signal indirectly by means of a sensing matrix (for example, the partial Fourier matrix in \citet{Candes:2007p2815}) $\psi$ in Eq. (\ref
{equ_Isparse}) or $\Theta$ in Eq. (\ref{equ_basissparse}). As proven in~\citet{Candes:2007p2815},  given a random selected measurements or observations $V$, the target signal $I$ can be recovered exactly with overwhelmingly probability, when the number of measurements $m$ satisfies
\begin{equation}\label{equ_incoherent}
m  \geq \; \mathrm{Const}.\: \mu^2(F, \phi)\cdot s\cdot \mathrm{log}\;{n},
\end{equation}where $s$ is the number of non-zero entries in $\alpha$, $\mu(F, \phi)$ is the mutual coherence, and $F$ is the sensing system. Note that, $\psi$ is a subset of $F$.
For example, if the sensing system is in the Fourier domain, then the partial Fourier matrix $\psi$ is assembled from the randomly selected rows of the Fourier transform matrix $F$. The unit-normalised basis vectors of $F$ are organized in rows, and the unit-normalised basis vectors of $\phi$ are organized in columns. CS requires that the correlation or coherence between $F$ and $\phi$ is low. This is measured by the mutual  $\mu(F, \phi)$
\begin{equation}
\label{equ_mutrualcoherence}
\mu(F, \phi)=\sqrt{n}\cdot\displaystyle{\mathop{\mbox{max}}_{1 \leq k, j \leq l}}|<F_k, \phi_j>|,
\end{equation} where $|<F_k, \phi_j>|$ is the inner product between two vectors: the $k^{th}$ row of $F$ and the $j^{th}$ column of $\phi$.
The coherence measures the largest correlation between any two rows of $F$ and $\phi$. If 
$F$ and $\phi$ contain highly correlated elements, the coherence is large, and vice-versa. 

Equation (\ref{equ_incoherent}) shows that as the coherence decreases, fewer samples are needed. 
Thus if there is a sparse representation in $\phi$, it must be spread out in the domain $F$ in which it is acquired. For example, a spike signal in the time domain can be spread out in the frequency domain. Therefore, the best way is to sense this signal in the frequency domain, because in this case $\mu(F, \phi)=1$. This makes perfect sense because to measure the strength of a single pulse, all that is needed is a single measurement in Fourier space, as opposed to an exhaustive search in the time domain.
 %\stackrel \overset
 
\subsection{Restricted isometry property (RIP)}
\label{sec:cs_rip} 

The discussion above focused only on the situation that the signal has a 
sparse representation with respect to a certain dictionary of basis functions. We note that ``sparse" here means there are few non-zero 
entries. However, this is rarely the case, and more often most of the entries will be approximately zero. CS theory is also applicable to this case but to explore it we need to introduce another concept, the restricted isometry 
property. In Eq. (\ref{equ_basissparse}), $\Theta$ is now the sensing matrix and $\alpha$ is a near 
sparse vector signal with  $s$ of the largest entries. For each integer $s=1,2,\cdots$, we define the isometry 
constant  $\zeta_s$ in the matrices $\Theta$ as the smallest number such that
\begin{equation}
\label{equ_rip}
(1-\zeta_s) \| \alpha \|^2_{l_2} \leq \| \Theta \alpha \|^2_{l_2}\leq (1+\zeta_s)\|\alpha\|^2_{l_2}
\end{equation} which holds for all $s$ sparse vectors $\alpha$. If $\zeta_s$ is not too close to 1, then an 
equivalent description of RIP is to say that all subsets of $s$ columns taken from the matrix $\Theta
$ are {\em approximately} orthogonal. It cannot, of course, be exactly orthogonal, because it has more columns than 
rows. \citet{Candes:2004p2832} shows that if the matrix $\Theta$ satisfies the RIP of order $2s
$ and the isometry constant $\zeta_{2s}<\sqrt{2}-1=0.414$, then the L1 norm minimization solution $\alpha^*$ to Eq. (\ref{equ_basissparse}) obeys:
\begin{equation}
\label{equ_nonoise_l1}
\| \alpha^*-\alpha\|_{l_2} \leq Const \cdot \| \alpha- \alpha_s \|_{l_1}/\sqrt{s},
\end{equation}where $\alpha_s$ is the vector $\alpha$ with all but the largest $s$ entries set to 
zero. 

In practical applications, the measured data will be corrupted by noise. To investigate the effects of noise on CS, Eq. (\ref{equ_basissparse}) can be rewritten as
\begin{equation}
\label{equ_noisesparse}
\Theta\alpha+E=V,
\end{equation} where $E$ is a stochastic error term. 
If the matrix $\Theta$ satisfies a RIP of order $2s$ and the isometry constant $\zeta_
{2s}<\sqrt{2}-1=0.414$, then the L1 norm minimization solution to Eq. (\ref{equ_noisesparse}) satisfies
\begin{equation}
\label{equ_noisel1}
\|\alpha^*- \alpha \|_{l_2} \leq C_0 \cdot \| \alpha-\alpha_s\|_{l_1}/ \sqrt{s}+C_1\cdot \epsilon,
\end{equation} where $C_0$ and $C_1$ are constants and $\epsilon$ bounds the amount of 
noise $E$ in L2 norm. 

When the RIP is satisfied, the sensing matrices are almost orthogonal. Examples of matrices that satisfy the RIP are the Gaussian sensing matrix, binary sensing matrix, random orthonormal sensing matrix, and partial Fourier sensing matrix~\citep{Candes:2006p23}. A Gaussian sensing matrix can be achieved by selecting $m$ rows randomly and independently from the normal distribution matrix of size $n$ by $n$ with mean zero and variance $1/m$.  A partial Fourier sensing matrix can be constructed by selecting $m$ rows randomly and renormalizing the columns to unity. These random sensing matrices are not only for $\Theta$ 
in Eq. (\ref{equ_basissparse}) but also for the matrix $\psi$, provided that $\phi$ is an arbitrary 
orthonormal basis~\citep{Candes:2008p14}. These random sensing matrices are 
universal sensing matrices in a certain sense, because randomness can always help in designing 
a suitable sensing matrix that obeys the RIP.
  
\subsection{Reconstruction}
\label{sec:cs_reconstruction} 
 
As described above, CS theory shows that  we only need measure a compressed version of the sparse target signal. However, a decompression or reconstruction procedure is obviously required to recover the signal. From equations (\ref
{equ_nonoise_l1}) and (\ref{equ_noisel1}), we can see that L1 norm minimization can give a good solution. The L1 norm of a vector is simply the sum of the absolute values of the
vector components.  L1 norm minimization has been studied well and its history can be traced back to geophysics in the mid-eighties~\citep{Santosa:1986p502}. 

Compressive sensing uses a basis pursuit (BP) approach to find the solution to Eq. (\ref
{equ_basissparse}) by solving
\begin{equation}
\label{equ_bpnonoise}
{\rm{min}}\; \| \alpha \|_{l_1} \; \; s.t.\; \psi\phi\alpha=V.
\end{equation} BP is a principle for decomposing a signal into an superposition of dictionary 
elements, in which these coefficients have the smallest  L1 norm among all possible 
decompositions. For measurements contaminated by noise, BP can be replaced by a more general algorithm: Basis Pursuit De-Noise (BPDN).~\citep{Chen:1998p2836}. The solution is given by
 \begin{equation}
\label{equ_bpnoise}
{\rm{min}}\; \| \alpha \|_{l_1} \; \; s.t. \; \|  \psi\phi\alpha-V \|_{l_2}  \leq \; {     \epsilon}.
\end{equation} 
Because the L1 
norm is convex, it can then be calculated by modern linear programming optimization 
algorithms~\citep{Beck:2009p710,Becker:2009p2860,Boyd:2004p3016}. There are many L1 norm 
solvers or toolboxes for L1 norm minimization problems. 
  
We, finally, note that CS is an asymmetric compression method, which means that the reconstruction step is unrelated to the measurement method. For general compression methods, the decompression step has to be the inverse of the compression step, for example, if we compress an image with a JPEG codec, then we have to decompress the image with a JPEG decoder. For compressive sensing, for the same measurements there are many ways to reconstruct or decompress the target signal. For example, by selecting the different decomposition transform matrix $\phi$ in equations~\ref {equ_bpnonoise} and~\ref{equ_bpnoise}, different reconstruction results will be calculated. We note that the best solution depends on an understanding of which domain the signal has the sparsest representation.

\section{Radio interferometry}
\label{sec:radiointro} 

A radio interferometer
(\textit{i.e.} a pair of antennas is used to measure the spatial coherence (visibility) function due to the sky brightness within the field of view of the antennas. 
The van Cittert-Zernike theorem states that the visibility  $V(u,v)$ is a two-dimensional Fourier transform of the
sky brightness $I$
\begin{equation}\label{equ_vancittert}
   V(u,v)=\int I(x,y) e^{2\pi j{(ux+vy)}}dxdy, 
\end{equation} where $u$ and $v$ are the baseline vectors of the interferometer and $I(x,y)$ is the 
sky brightness of the coordinates $x$ and $y$. Here, we assume that we can ignore the component of the baseline towards the source - an approximation that is
adequate for a small field of view. Equation (\ref{equ_vancittert}) shows that the radio telescope 
array measures the Fourier coefficients of the sky brightness. If we could measure all the Fourier 
coefficients of the sky brightness, the sky brightness image could be reconstructed using the 
inverse Fourier transform. Unfortunately, it is usually infeasible and certainly expensive to capture all the visibility data, up to the Nyquist-Shannon sampling limit. Thus, the measurements are the partial and incomplete Fourier coefficients. This situation can be described as
\begin{equation}\label{equ_mask}
     MFI=V,
\end{equation} where $I$ is the sky brightness image in a vector format of length $n$, and $V$ is the measured visibility data $i.e.$ the Fourier coefficients in a vector format of length $n$ as well. Strictly speaking, this is not quite consistent with the previous definition of $V$ as it now contains many zeros. We note that these unmeasured visibility data in $V$ are denoted here as zeros, where $F$ is the Fourier transform matrix of size $n\times n$ and $M$ is a binary matrix implementing the UV coverage mask of size $n\times n$. In $M$, it contains either zeros on each line at the index of the unmeasurable visibility data, or only one non-zero value on each line at the index of measurable visibility data. Since we do not capture all the visibility data using a telescope array, there are many zero lines in the mask matrix $M$.

Radio interferometric imaging can be described by Eq. (\ref{equ_mask}). If we carry 
out an inverse Fourier transformation on both sides, we have
\begin{equation}\label{equ_Fmask}
     F^{-1}M\ast I=F^{-1}V,
   \end{equation}  where $\ast$ is a convolution operator and $F^{-1}$ is the inverse Fourier transform matrix. 
   Here, $F^{-1}M$ is called the point spread function or the dirty beam. From Eq.(\ref{equ_mask}), we know that many entries of  $V$ are zero. If we apply an inverse Fourier transform to this, we will get $F^{-1}V$, which is also called the dirty map, $i.e.$, the convolution of the true brightness distribution image $I$ with the dirty beam $F^{-1}M$. Therefore, by long-established convention, the reconstruction of $I$ is called deconvolution. 

   In our investigations described below, we consider two deconvolution methods: the commonly used H\"ogbom CLEAN~\citep{Hogbom:1974p1206} and the multiscale CLEAN~\citep{Cornwell:2008p191}. The H\"ogbom CLEAN method is a matching pursuit algorithm. It repeats the two main steps: finding the peak value in the residual image and removing the dirty beam at that position with a certain gain to obtain an upgraded residual image for the next iteration. CLEAN is a non-linear method and it works well for point-like sources. Multiscale CLEAN is a scale-sensitive version of the CLEAN method designed to reconstruct extended objects~\citep{Cornwell:2008p191}.

This experiment is based on simulating the Australian Square Kilometer Array Pathfinder (ASKAP)~\citep{2009IEEEP..97.1507D} radio telescope, which comprises an array of 36 antennas each 12m in diameter. When completed in 2013, ASKAP will provide high dynamic range imaging with wide-field-of-view phased array feeds.  The 36 antennas will be distributed with a smallest separation of 22 m and a longest baseline 6km in Boolardy of Western Australia. Moreover, 30 of them will be located within a 2km radius core with an approximately Gaussian baseline distribution. The Gaussian baseline distribution measures large magnitude samples as discussed above, and produces a Gaussian shape point spread function (PSF) in the image domain, which can be approximated well by a clean beam and is well-behaved when deconvolution is not needed. In ASKAP, there are also six long baseline telescopes to capture high frequency detail, such as the separation of compact sources. The design of ASKAP is well-adapted to CS, although it was designed independently of CS. Although ASKAP is equipped with phased array feeds to give approximately 30 primary beams on the sky, in our simulations we consider only one primary beam. We expect that the results will transfer over to the case of multiple primary beams, though this remains to be demonstrated in a subsequent paper.

\section{Compressive sensing-based deconvolution methods}
\label{sec:csdeconvolution} 
We now briefly discuss the conventional understanding of the deconvolution problem in radio interferometry. As described above, the limited sampling of the Fourier plane means that there will be many solutions to the convolution equation. It is therefore necessary to use some prior information to select a solution. The prior information can be that
the true sky brightness is real and positive, the sky brightness image has only point sources that 
are sparse in the image domain, or the sky brightness image has a sparse presentation with respect to a dictionary of basis functions, to name just a few. In some other imaging contexts, regularisation methods that minimize the L2 norm solutions are used, but the application of these techniques is unsuitable for deconvolution problems in radio astronomy. The L2 norm minimization of Eq. (\ref{equ_mask}) provides the dirty map, which is the convolution of the true brightness distribution with the dirty beam. Compared to the L2 norm, the L1 norm used in CS theory puts more weight on the small magnitude variables. This means that L1-minimizing solutions will have a larger number of small values. We now see the connection to CS.

Returning now to CS, we see that CS can be applied to radio astronomy in a straightforward manner, because the Fourier domain is perfectly incoherent to the image domain~\citep{Candes:2004p2832}.  Equation (\ref{equ_mask}) can also be rewritten in a similar form to Eq. (\ref{equ_Isparse}), when $MF=\psi$. This allows the reconstruction of point sources and compact sources, these being sparse in the image domain. However, it will not work effectively for some extended sources such as galaxies, gas emission, and nebulae. Fortunately, in these cases, a sparse representation of the signal can be achieved by using some dictionary functions such as the wavelet transform and the undecimated wavelet transform~\citep{Starck:2007p1136}.  

Finally we wish to emphasize the role of the sampling matrix.
Even though we know that the noiselet~\citep{Coifman200127} and the Haar wavelet transforms~\citep{chui1992introduction} are a perfect incoherent pair~\citep{Candes:2007p2815}, the radio interferometer can only measure the visibility data $i.e.$ the Fourier coefficients. That is, we are limited to the Fourier domain for the sampling procedure. A hypothetical instrument might measure in one or the other of noiselet or Haar wavelet space, in which case the other space could be used for
the signal.
  
In our studies below, we test the following two CS-based methods for radio astronomy deconvolution: partial Fourier and Isotropic undecimated wavelet transform based CS. 

\subsection{Partial Fourier (PF) based CS deconvolution method}
\label{subsec:pf} 

In this paper, PF is an
abbreviation for an L1 norm based partial Fourier reconstruction method. The partial Fourier sensing 
matrix exactly matches the radio astronomy array described above.  We can apply the L1-norm-based 
method in a straightforward manner by rewriting the deconvolution problem as
\begin{equation}
\label{equ_pfnonoise}
{\rm{min}}\; \| I \|_{l_1}   \; \;\; s.t.\; MFI=V,
\end{equation} for the pure deconvolution problem without noise,

\begin{equation}
\label{equ_pfnoise}
{\rm{min}}\; \| I \|_{l_1}  \;\; s.t. \; \| MFI-V \|_{l_2}  \leq \; {     \epsilon},
\end{equation} for the noise contaminated cases, where $\epsilon$ describes the uncertainty in 
the observation $V$ as in the situation where the measurements are contaminated with noise. We solve the 
above equation by using the fast iterative shrinkage-thresholding algorithm~\citep{Beck:2009p710} described later.

\subsection{Isotropic undecimated wavelet transform (IUWT) based CS}
\label{subsec:iuwt} 

The PF can work effectively when the sky brightness images include point 
sources. In many circumstances, this is not the case. However, as long as we can find a suitable dictionary of basis functions in which
the extended sources possess a sparse representation, the deconvolution procedure can still be carried out.
In this paper, we adopt the isotropic undecimated wavelet transform (IUWT)~\citep{Starck:2007p1136} 
as the dictionary $\phi$. The IUWT algorithm is very well suited to astronomical images, because both the UDWT and the RDWT (undecimated/ redundant  wavelet transform) preserve translation-invariance, and many sources in the universe are isotropic~\citep{Starck:2006p813}. In the IUWT, a non-orthogonal filter bank is used. The low pass filter is $h^{1D}=[1,4,6,4,1]/16$, and the high pass filter $g^{1D}=\delta-h^{1D}=[-1,-4,-10,-4,-1]/16$). In the implementation of the IUWT, we need only apply the low pass filter~\citep{Starck:2007p1136}. The high frequency components for the next scale can be calculated by subtracting the low frequency components from the current scale. 

If the number of scales of the wavelet transform is $l$, then the IUWT has $l+1$ images; in contrast, there are $3 l +1$ times as many sub-band images with the UDWT or the RDWT. Therefore, it requires less computational time and less memory than the UDWT/RDWT.

If we denote the isotropic undecimated wavelet transform as $W$ and its inverse transform of IUWT as $W^{-1}$, then the IUWT-based CS deconvolution method can be written as  

\begin{equation}
\label{equ_l1magic}
{\rm{min}} \;\| \alpha \|_{l_1} \; \; s.t.   \; MF W^{-1} \alpha=V,
\end{equation} where $I=W^{-1}\alpha$ and $\alpha$ denotes the wavelet coefficients in the IUWT domain in a vector format. Here, $\alpha$ is a sparse representation of the true sky brightness $I$ in which there are expanded sources. For the noise contaminated case, we have

\begin{equation}
\label{equ_l1magicnoise}
{\rm{min}} \;\| \alpha \|_{l_1}  \; \;s.t. \;  \| MF W^{-1} \alpha-V \|_{l_2} \leq \epsilon.
\end{equation}

The two above equations can also be rewritten in Lagrangian form
\begin{equation} \label{equ_L1minfisti}
    {{\rm{min}}} \;\; \lambda | \alpha |+ \| MFW^{-1}\alpha-V \|^2.
\end{equation} From a Bayesian point of view, $\lambda$ is a balancing parameter between the 
contribution from the maximum likelihood part and the prior part $i.e.$ the second term and the 
first term, respectively. There are many L1-norm-minimization algorithms or toolboxes for these problems, such as, L1-magic, which is a second-order gradient-based optimization method, and some 
first-order gradient methods such as the iterative methods ISTA (Iterative Soft-Thresholding Algorithm~\citep
{Daubechies:2003p1207}) and FISTA (Fast Iterative Shrinkage-Thresholding Algorithm~\citep{Beck:2009p710}). For large-scale problems, first-order methods are preferable, because the 
calculation of the inverse of the Hessian matrices for the second order methods is slow~\citep{Beck:2009p710}.  

Using the commonly-adopted first-order gradient method ISTA~\citep
{Daubechies:2003p1207}, the solution can be calculated by
\begin{equation}
\label{ista_solution }
\alpha_{k+1}=\mathcal{T}_{\lambda t}(\alpha_{k}- 2t(MFW^{-1})^T(MFW^{-1}\alpha_{k}-V)),      
\end{equation}where $t$ is an appropriate step size and $\mathcal{T}_{x}$ denotes the shrinkage operator for each component $\alpha_{ki}$ in $\alpha_{k}$ of the $k^{th}$ iteration as
\begin{equation}
\label{ shrinkage}
\mathcal{T}_{x}=(|\alpha_{ki}|-x)_{+}sgn(\alpha_{ki})  ,  \;\;  \forall \alpha_{ki}\in \alpha_{k} .
\end{equation}
The above equation can be rewritten as
\begin{equation}
\label{ista_further}
\alpha_{k+1}=\mathcal{T}_{\lambda t}(\alpha_{k}- 2tWF^{-1}M^{T}(MFW^{-1}\alpha_{k}-V)),
\end{equation} because both matrices $W$ and $F$ are orthogonal. The initial $\alpha_{0}$ can be calculated by transforming the dirtymap into the IUWT domain. After a certain number of iterations, the wavelet coefficients of the reconstructed image can be calculated from Eq. (\ref{ista_further}). We then apply the inverse wavelet transform to those wavelet coefficients, the model image $I$ eventually reconstructed.

However, ISTA converges quite slowly. \citet{Beck:2009p710} developed a more rapidly convergent variant: Fast ISTA (FISTA). In this approach, the iterative shrinkage operator is not employed on the previous point $\alpha_{k}$, but rather at a very specific linear combination of the previous two points $\alpha_{k}$ and $\alpha_{k-1}$.  Based on FISTA, the solution to Eq. (\ref{equ_L1minfisti}) is
\begin{equation}
\label{fista_solution }
\alpha_{k+1}=\mathcal{T}_{\lambda/L}(\beta_{k}- \frac{2}{L}(MFW^{-1})^T(MFW^{-1}\beta_{k}-V)),      
\end{equation}where $L$ is the Lipschitz constant in~\citet{Beck:2009p710} and $\mathcal{T}_{{\lambda}/{L}}$ denotes the shrinkage operator for each component $\alpha_{ki}$ in $\alpha_{k}$ of the $k^{th}$ iteration as
\begin{equation}
\label{ fistashrinkage}
\mathcal{T}_{{\lambda}/{L}}=(|\beta_{ki}|-\frac{\lambda}{L})_{+}sgn(\beta_{ki})  ,  \;\;  \forall \beta_{ki}\in \beta_{k},
\end{equation}

\begin{equation}
\label{Fistastep}
t_{k+1}=\frac{1+\sqrt{1+4t_k^2}}{2},
\end{equation}
\begin{equation}
\label{fistaconbined}
\beta_{k+1}=\alpha_k+(\frac{t_k-1}{t_{k+1}})(\alpha_k-\alpha_{k-1}). 
\end{equation} Both $\alpha_{0}$ and $\beta_{0}$ are set to be the wavelet coefficients of  the dirtymap in the IUWT domain. The initial $t$ is set to be 1 as suggested in~\citet{Beck:2009p710}. The algorithm  ends either after an assigned certain number of iterations or when the minimum of Eq. (\ref{equ_L1minfisti}) is attained.
After executing the algorithm, the wavelet coefficients of the reconstructed image can be calculated. The main computational effort in FISTA remains the same as in ISTA, namely, in the shrinkage operator. However, as
described in~\citet{Beck:2009p710}, for ISTA, the error decreases as $1/k$; for FISTA, the error decreases as $1/{k^2}$, where $k$ is the number of iterations. 

In this approach, the IUWT-based CS with FISTA, is abbreviated as IUWT-based CS in this paper. These CS-based deconvolution methods, PF and  IUWT-based CS,  are implemented in MATLAB. Our code may be found at: http://code.google.com/p/csra/downloads\footnotetext[1]{Download the file ``PF\_IUWT.zip" which includes both PF and IUWT-based CS algorithms}.

\section{Experimental results }
\label{sec:results} 

We compare the CS-based deconvolution methods (PF and IUWT-based CS) introduced above and the CLEAN-based approaches: H\"ogbom CLEAN and the multiscale CLEAN.

\begin{figure*}
 \begin{center}
      \begin{tabular}{cccc}
{     
       \subfigure  {    \includegraphics[scale=0.24]{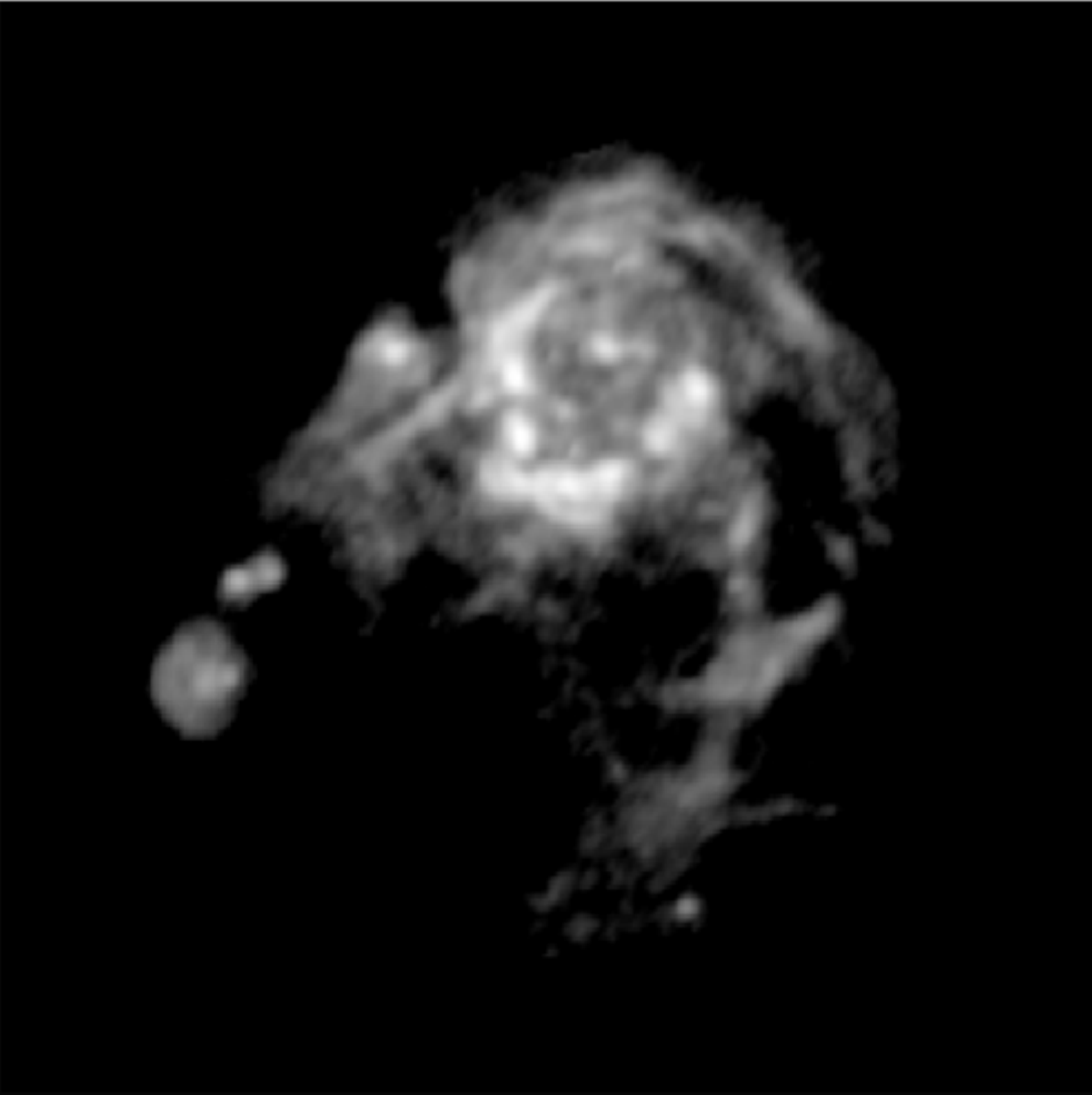}  \label{pre_m31zoom} }
       \subfigure  {    \includegraphics[scale=0.3]{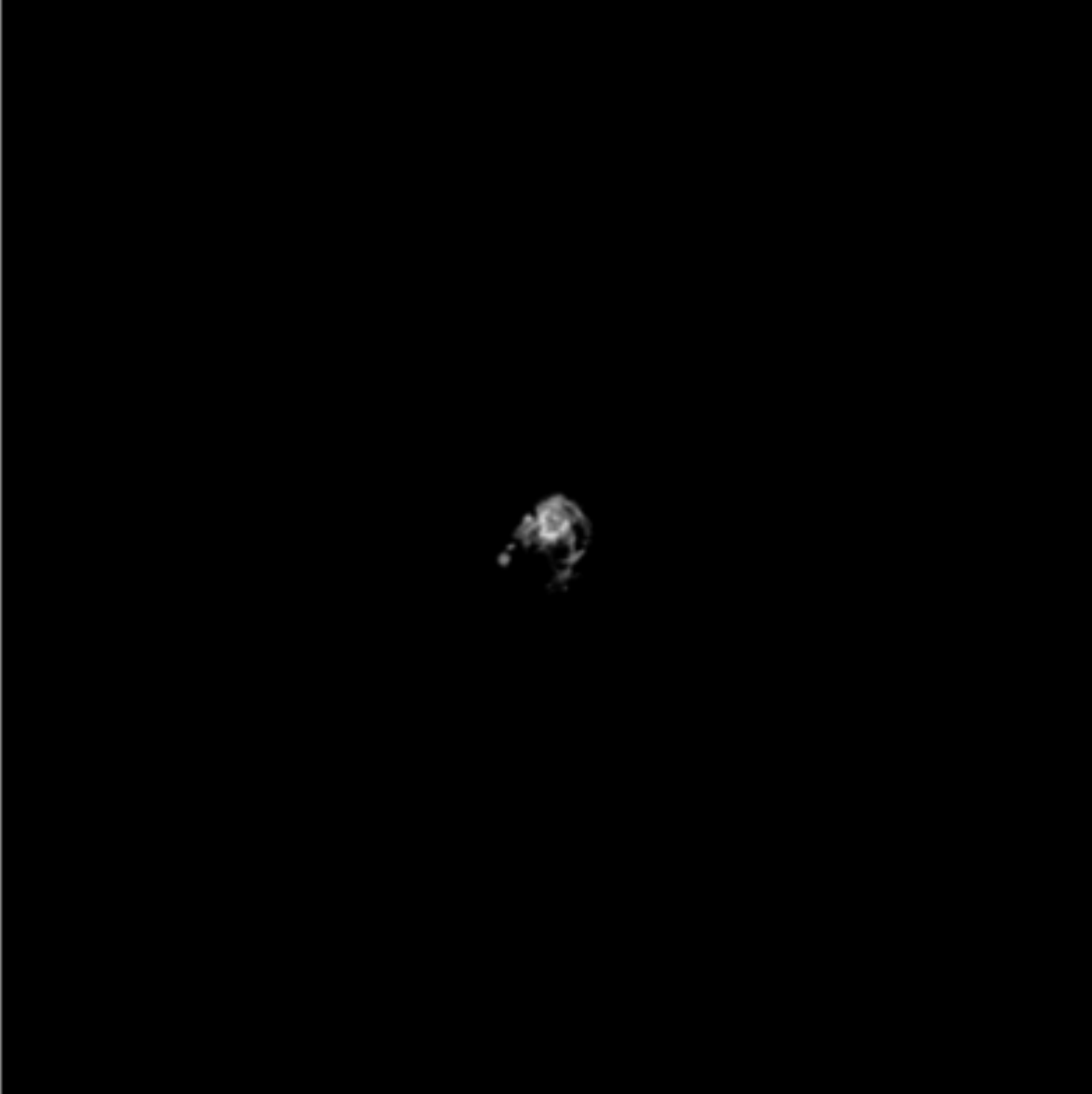} \label{pre_m31}}
       \subfigure  {    \includegraphics[scale=0.3]{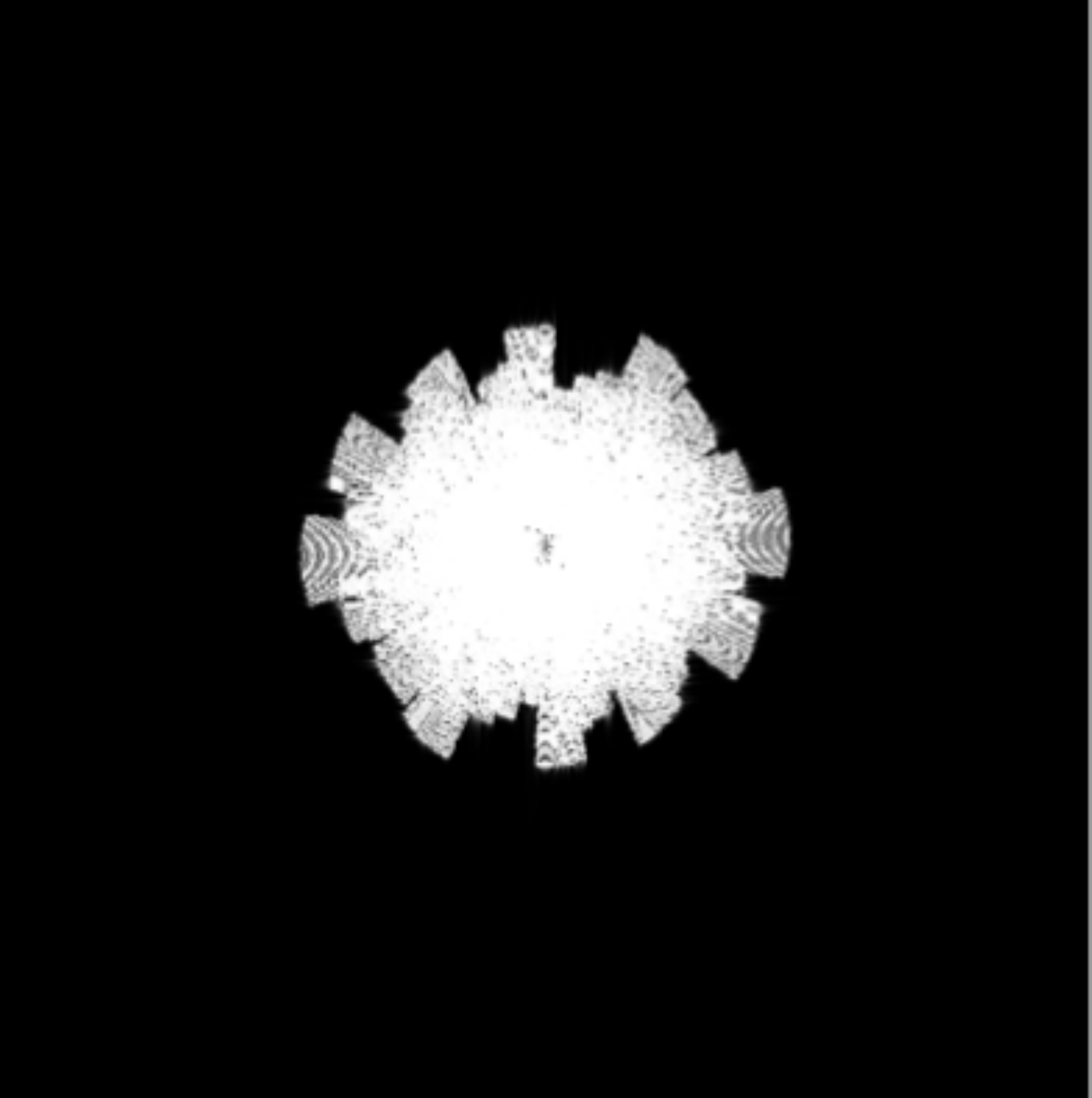} \label{pre_uvwiener}}
       \subfigure  {    \includegraphics[scale=0.3]{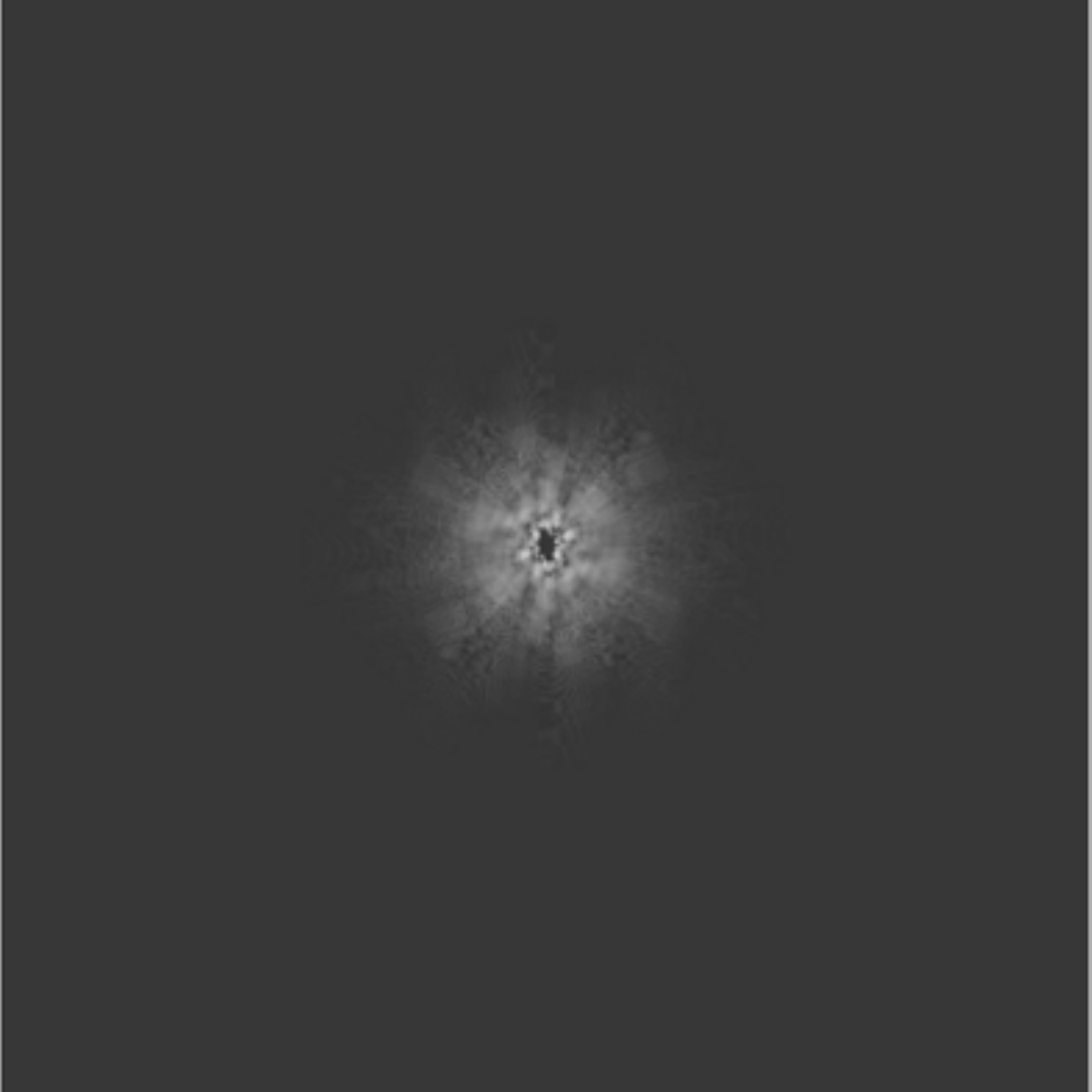} \label{pre_uvnone}}
}
 \end{tabular}
  \begin{tabular}{cccc}
{
      \subfigure  {   \includegraphics[scale=0.3]{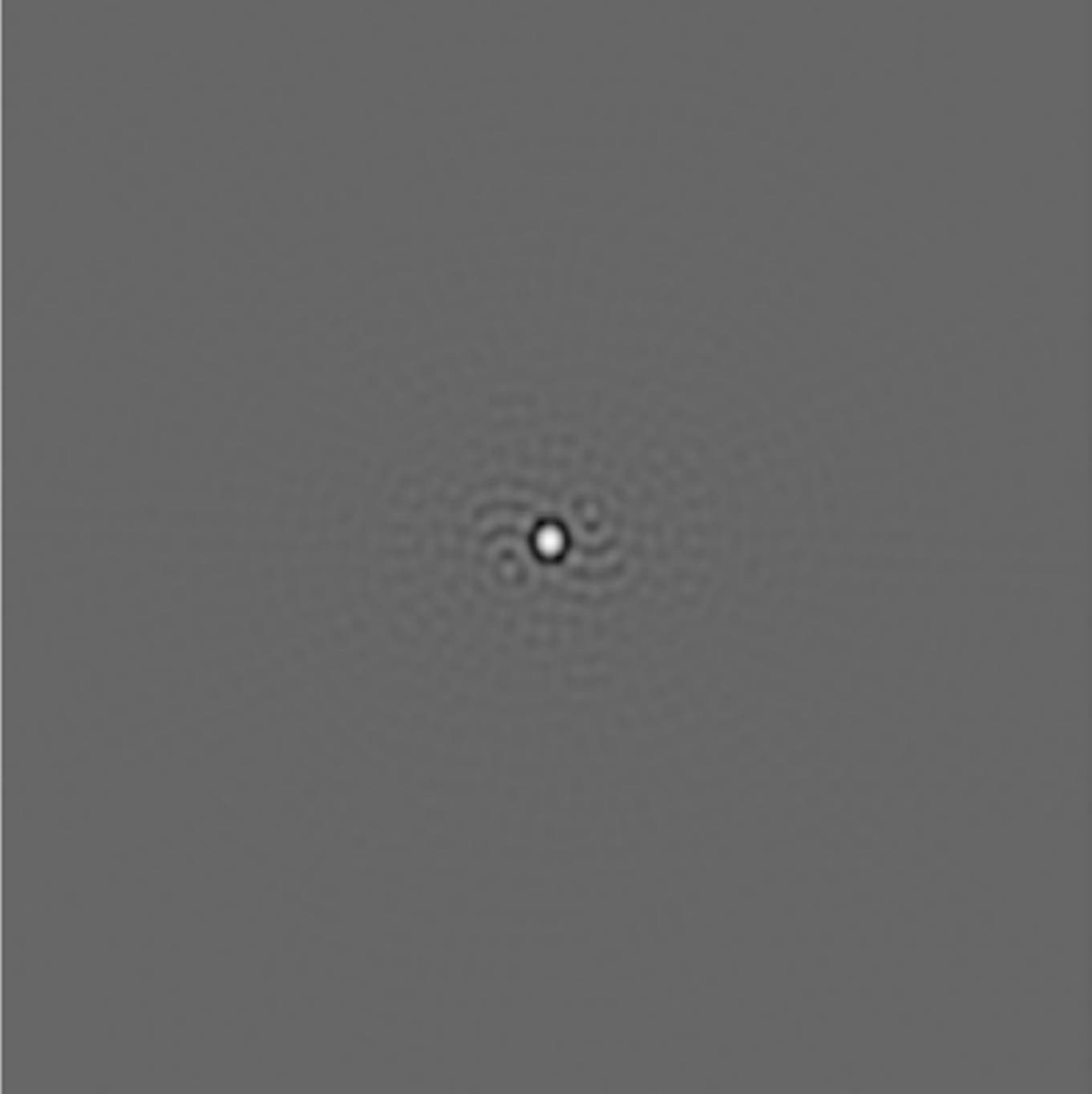} \label{pre_psfwiener}}
       \subfigure {   \includegraphics[scale=0.3]{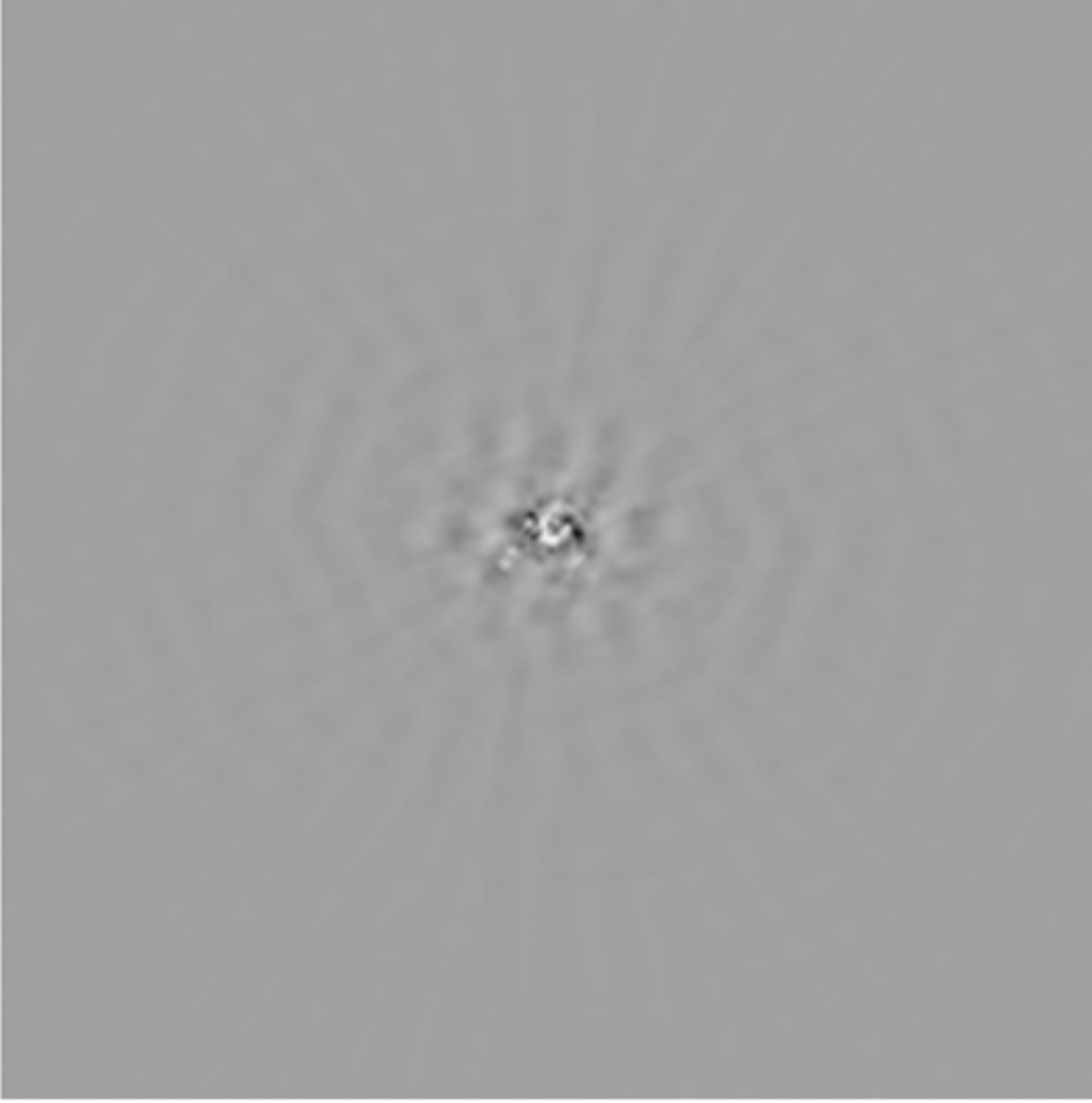} \label{pre_psfnone} }
       \subfigure  {  \includegraphics[scale=0.3]{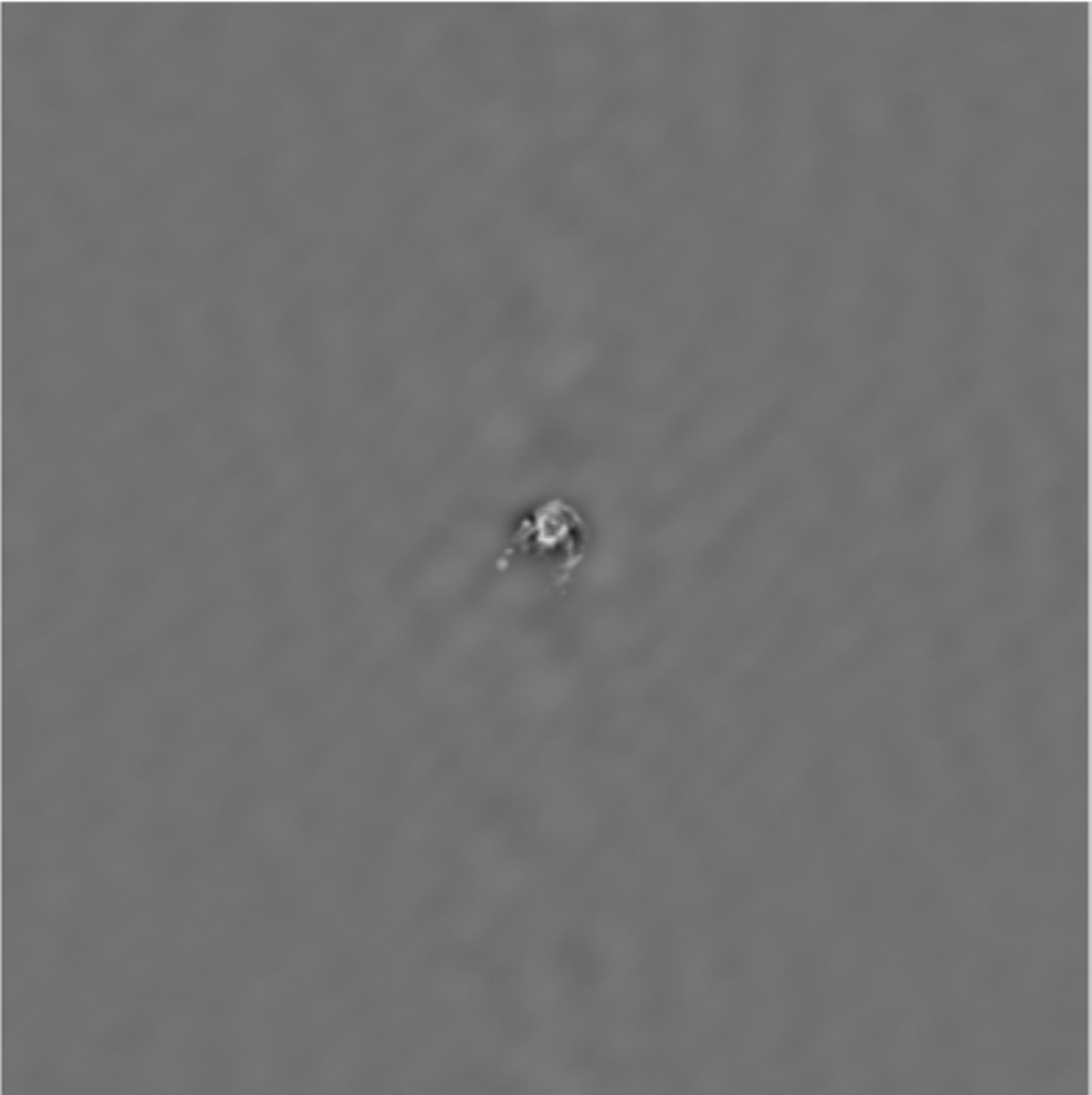} \label{pre_dirtywiener}}
       \subfigure  {   \includegraphics[scale=0.3]{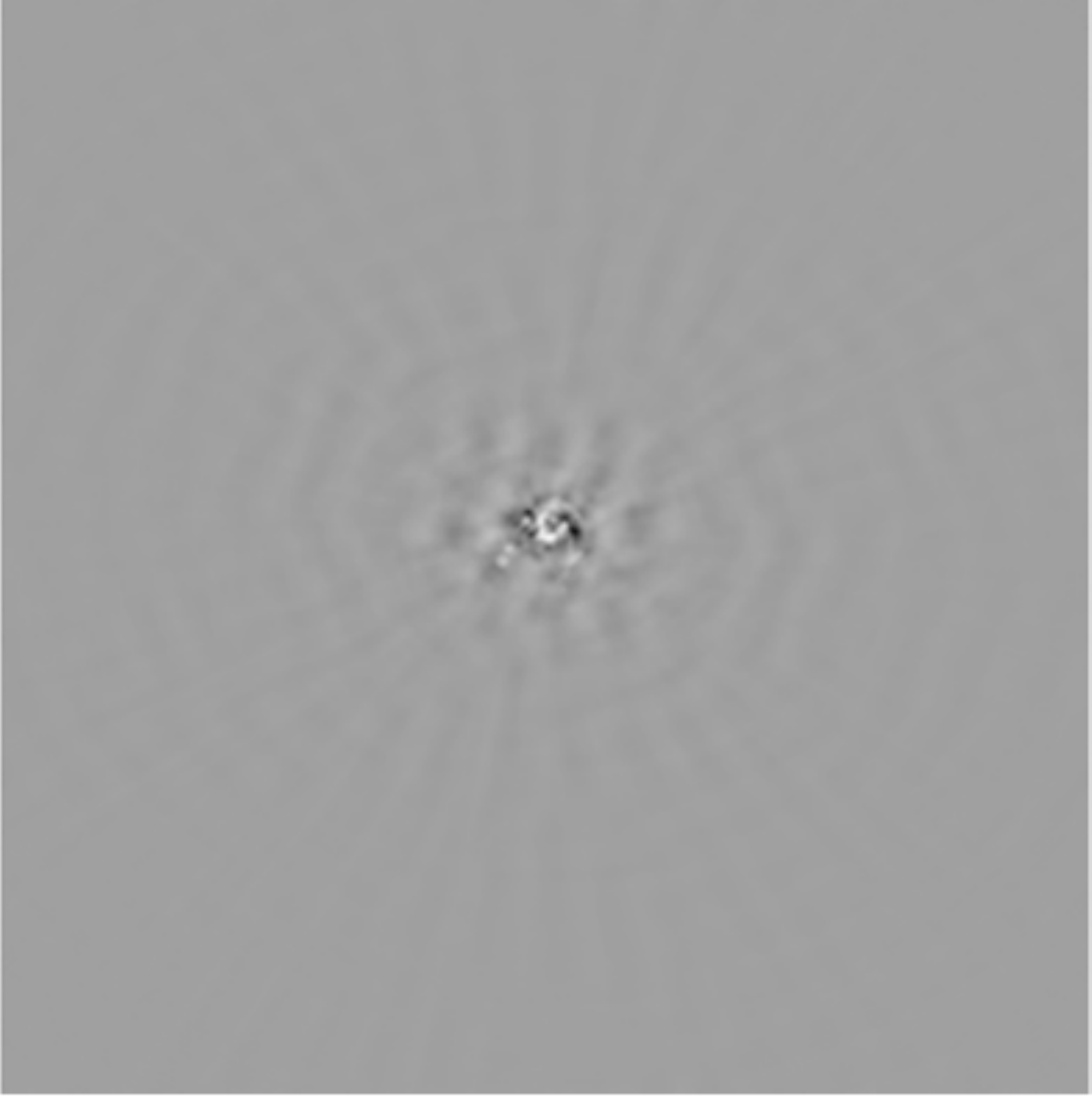} \label{pre_dirtynone}}
 }
 \end{tabular}
 
  \end{center}
\caption{Testing data shown with the scaling power of $-1.5$. From left to right in the first row are: ~\subref{pre_m31zoom} the test image  of size $256 \times 256$, \subref{pre_m31} the zero-padded true-sky image of size $2048 \times 2048$, \subref{pre_uvwiener}  the uniform-weighted UV coverage of ASKAP,  \subref{pre_uvnone} natural-weighted UV coverage of ASKAP; From left to right in the second row are:
\subref{pre_psfwiener} the corresponding PSF of the uniform-weighted UV coverage in a three-times-enlarged version, \subref{pre_psfnone} the corresponding PSF of natural-weighted UV coverage in a three-times-enlarged version, \subref{pre_dirtywiener} the dirtymap with the uniform-weighted UV coverage, and \subref{pre_dirtynone} the dirtymap with the natural-weighted UV coverage. } \label{fig_pr}
\end{figure*}

Our test image, of size $256\times 256$ pixels, is shown in the top left image of figure~\ref{fig_pr}. The brightness of the test image ranges from 0 to 0.0065Jy/pixel, and its total flux is 10Jy. The primary beam of ASKAP is 1.43 degrees for the working frequency of 1GHz. We only select the centre 30 antennas in this test, hence the highest achievable angular resolution is about 30 arcsec. To obey the Nyquist-Shannon sampling requirements, we adopt a cell size of 6 arcsec. The image size of ASKAP was selected to be $2048\times 2048$ pixels in this paper. The test image was padded with zeros in order to fit the simulation, and the zero-padded version of the true sky image is shown in figure~\ref{pre_m31}.

For the ASKAP antenna configuration, we obtain two UV coverages by using both uniform weighting and natural weighting, respectively. Uniform weighting can help us to control the point spread function and minimize the large positive or negative sidelobes. Natural weighting simply inversely weights all measurements with their variances thus optimises the signal-to-noise ratio in the dirty image. For most arrays, natural weighting creates a poor beam shape, because the measurements from shorter baselines are overemphasized~\citep
{Thompson:1986p2482}. In contrast, uniform weighting introduces a weighting factor that is the inverse of the areal density of the data in the UV plane. As a result, it can minimize the sidelobes and sharpen the beam at the expense of poorer sensitivity. Uniform weighting is more 
compatible with CS theory in that the UV coverage mask $M$ is a a binary mask that is identical to the derivation in section \ref{subsec:iuwt}. 

For natural weighting, however, the UV coverage mask $M$ is no longer a binary mask, and it ranges between 0 and the maximum value on the regular grid. In this case,
we cut the UV coverage mask $M$ by setting a threshold. The magnitude of $M$ smaller than the threshold will be set to zero by force. After this step, we can apply the above derivation in section \ref{subsec:iuwt} in a straightforward manner. The selection of the threshold for $M$ depends on the noise level. If there is no noise involved, the threshold can be arbitrarily small. However, this is not the case in reality. By our experience, the threshold between $0.1* max(M)$ and $0.01* max(M)$ is a good selection for ASKAP. Here, the threshold is set to be $0.01* max(M)$, i.e. any magnitude smaller than one percent of the maximum of $M$ will be set to zero. 

The source is taken to be located at declination -45 degrees and right ascension $12h30m00.00$ (epoch $J2000$), and the array is located at latitude -27degrees. The observing frequency range is between 700MHz and 1GHz. There are 30 channels with 10MHz bandwidth for each channel. This multifrequency observation will help us to fill the UV coverage. The integration time is 60 seconds, and the observing time is 1 hour. The system temperature controlling the noise level is set to 50K in this test. The two UV coverages with both uniform weighting and natural weighting can be seen in Figs.~\ref{pre_uvwiener} and~\subref{pre_uvnone}, respectively. In these UV coverages, the white dots show the visibility data sampled by the ASKAP array. The relevant dirty beams (normalised to unit peak) are shown in Figs.~\ref{pre_psfwiener} and \subref{pre_psfnone}, respectively. The dirty maps of uniform weighting and natural weighting can be seen in Figs.~\ref{pre_dirtywiener} and \subref{pre_dirtynone} in the same order. These dirty beams are displayed with an enlarged version to help readers identify the difference between these two dirty beams. We can see that uniform weighting produces a narrower beam than natural weighting. This difference can also be seen in the dirty maps.

\begin{figure*}
 \begin{center}
        \includegraphics[scale=0.5]{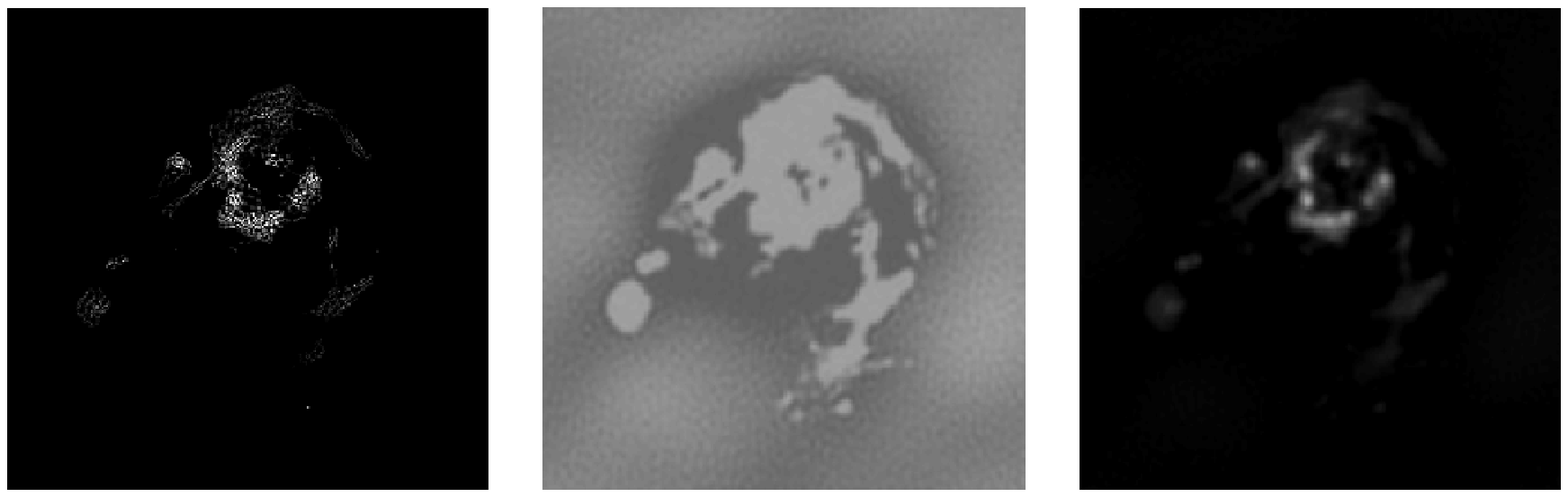}
         \includegraphics[scale=0.5]{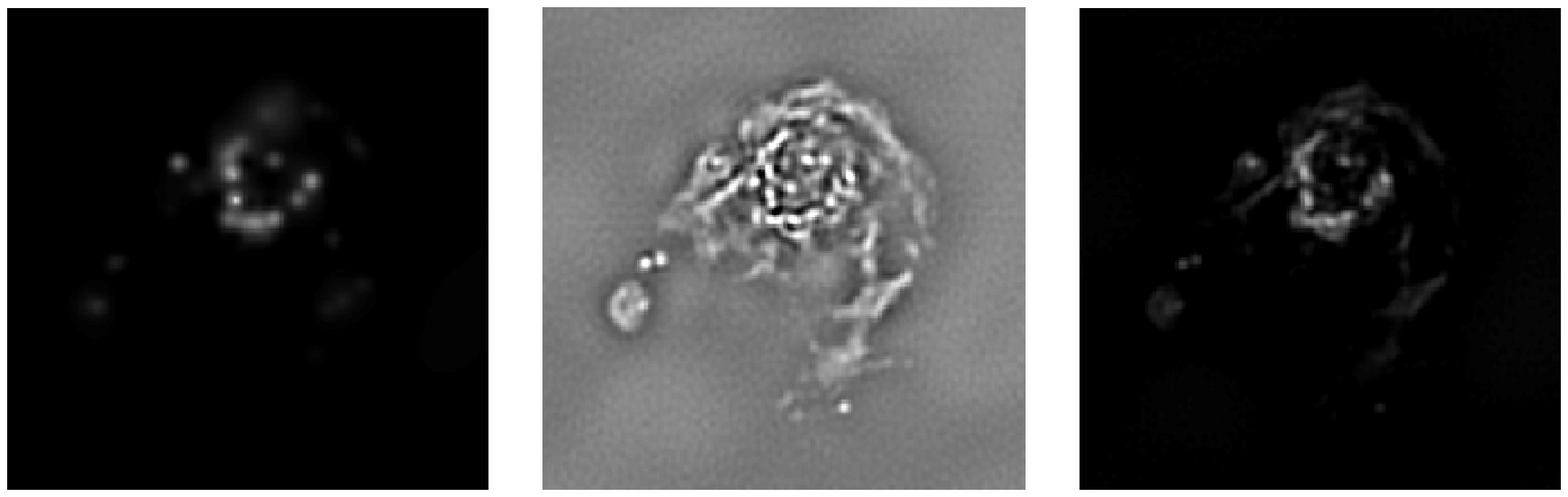}
         \includegraphics[scale=0.5]{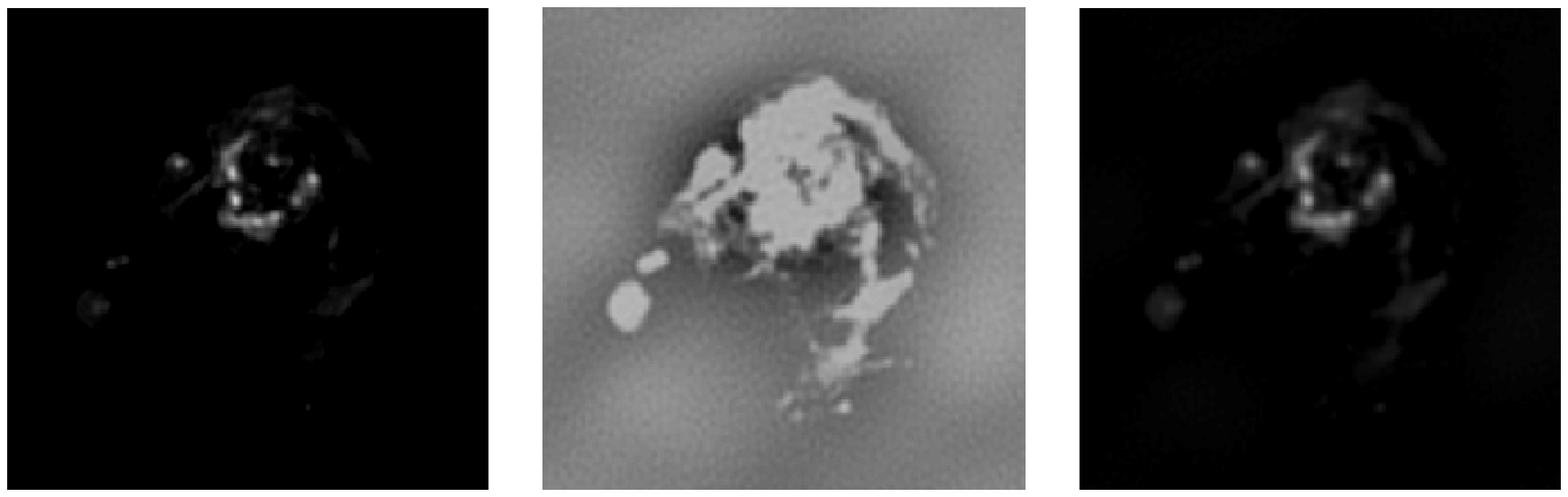}
        \includegraphics[scale=0.5]{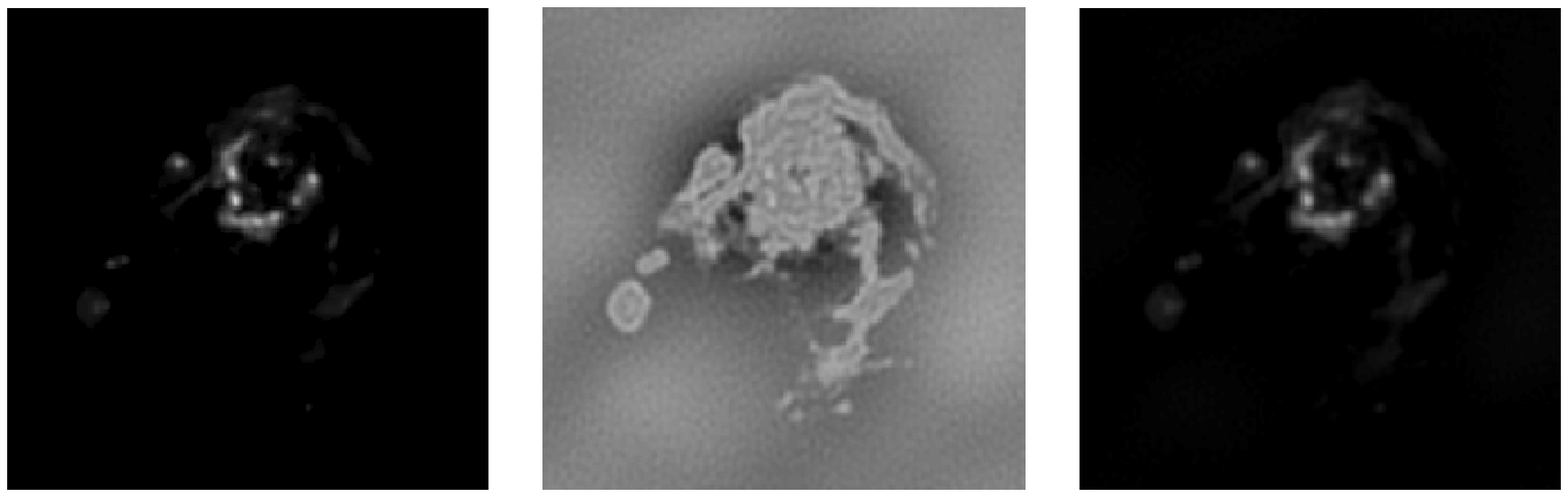}
 
 \end{center}
\caption{Deconvolution results with uniform weighting. From left to right in columns are the model images (shown with range 0 Jy/pixel to 0.006Jy/pixel and the scaling power of $-1.5$), the 
residual images (shown with range -0.01 Jy/pixel to 0.01 Jy/pixel and the scaling power of $0$), and the restored images (shown with range 0 Jy/pixel to 0.3 Jy/pixel and the scaling power of $-1.5$), respectively. From top to bottom in rows are the results of the H\"ogbom CLEAN, the multiscale CLEAN, the partial Fourier (PF) method, and the IUWT-based CS, respectively. } \label{fig_wiener}
\end{figure*}

\begin{figure*}
 \begin{center}
        \includegraphics[scale=0.5]{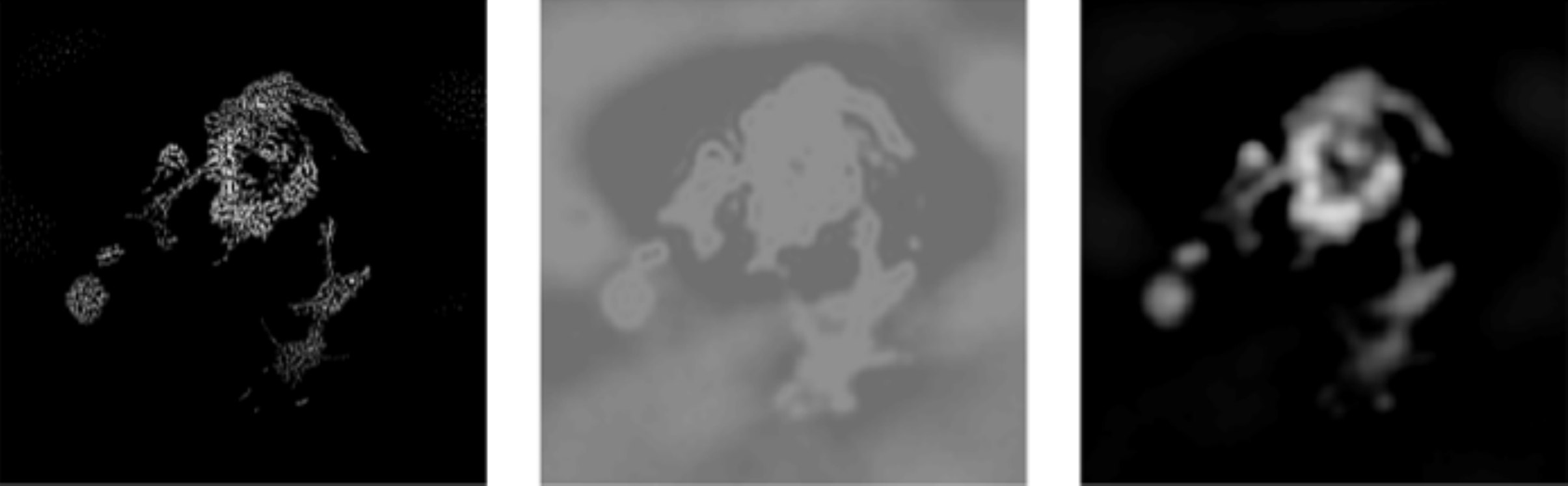}
         \includegraphics[scale=0.5]{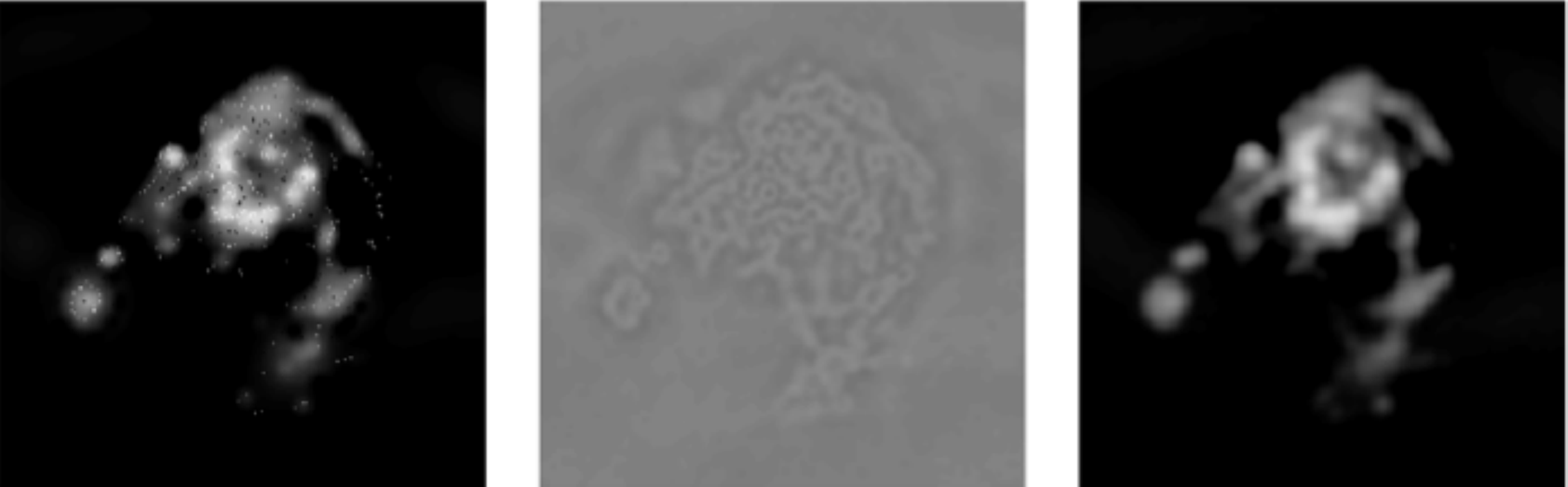}
         \includegraphics[scale=0.5]{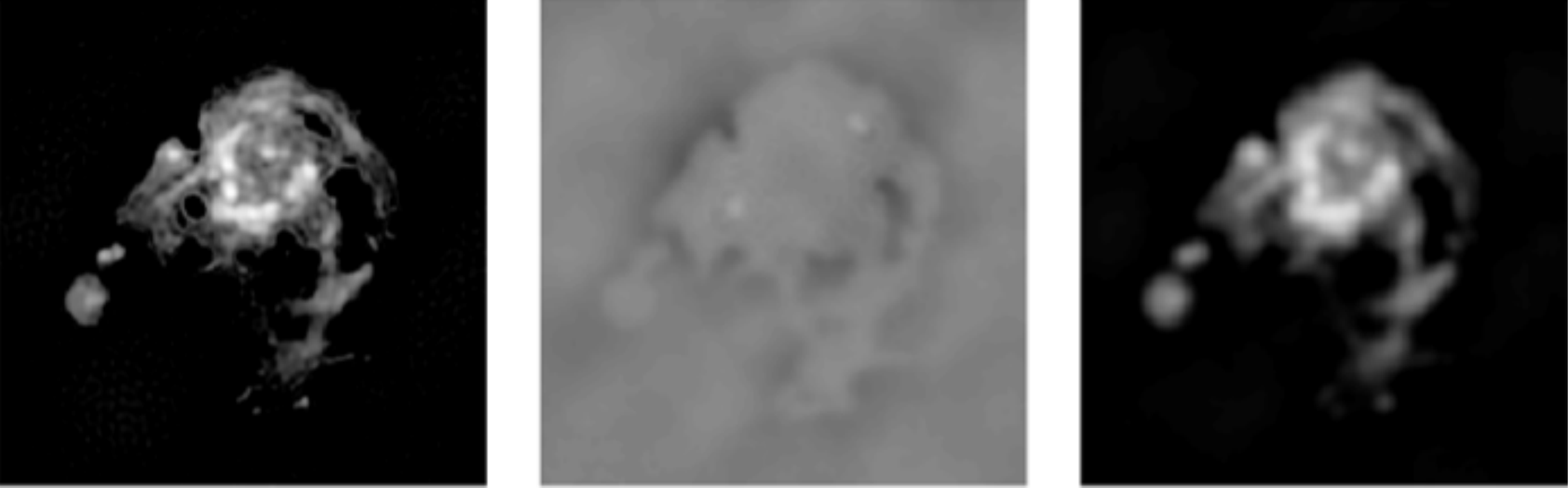}
        \includegraphics[scale=0.5]{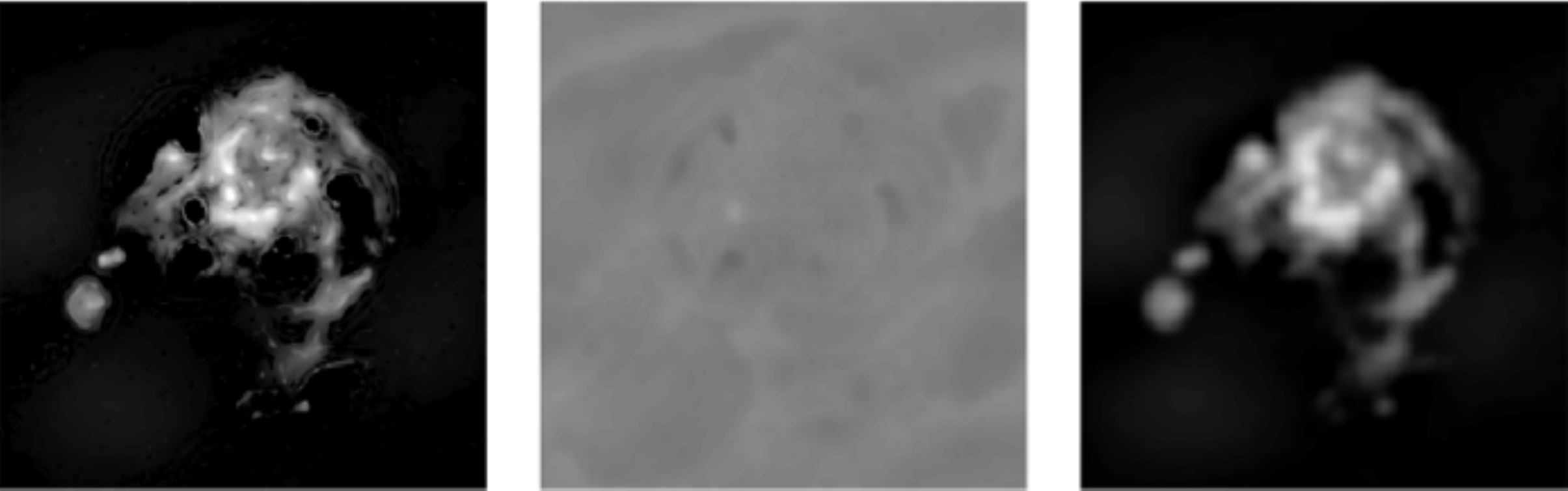}

 \end{center}
\caption{Deconvolution results with natural weighting. From left to right in columns are the model images (shown with range 0 Jy/pixel to 0.006Jy/pixel and the scaling power of $-1.5$), the 
residual images (shown with range -0.01 Jy/pixel to 0.01 Jy/pixel and the scaling power of $0$) and the restored images  (shown with range 0 Jy/pixel to 0.3 Jy/pixel and the scaling power of $-1.5$), respectively. From top to bottom in rows are the results of the H\"ogbom CLEAN, the multiscale CLEAN, the partial Fourier (PF) method and the IUWT-based CS, respectively. } \label{fig_none}
\end{figure*}

For the H\"ogbom CLEAN method, we use a 10000 iterations, a gain of 0.1, and a 0.001Jy threshold. For
the multiscale CLEAN method, the parameters are 10000 iterations, 
a gain of 0.7, six scales of  0, 2, 4, 8, 16, and 32, and a 0.001Jy threshold. For the CS based methods, $\lambda$ is set to $0.0001$ for both the PF and the IUWT-based CS.  It only takes 4 iterations for these CS based methods to converge i.e. the minimum of Eq. (\ref{equ_L1minfisti}) to be achieved. The clean beam is selected by fitting a two-dimensional Gaussian to the
dirty beam: 24.58 by 21.79 (arcsec), which is the Full Width at Half Maximum (FWHM). 
Since the cell size is 6 arcsec in this case,  the FWHM is 4.10 by 3.63 
pixels. We can then calculate a clean beam with standard deviation of 1.74 by 1.54. By using the clean beam, the relevant residual images and restored images can be computed by 
MATLAB. The residual image is defined to be
 \begin{eqnarray} \label{equ_residual}
 residual\; image  =&  dirty\; map- \nonumber \\
      & dirty\; beam * model,
\end{eqnarray} where $*$ denotes the convolution operator. The restored image is defined
\begin{eqnarray} \label{equ_restored}
 restored\; image  =& clean \; beam * model +\nonumber \\
      &residual \; image. 
\end{eqnarray}  
The deconvolved results with uniform weighting are shown in figure~\ref{fig_wiener}. From left to right in columns are the model images, the 
residual images, and the restored images, respectively. From top to bottom in rows are the results of the H\"ogbom CLEAN, the multiscale CLEAN, the partial Fourier (PF) method and the IUWT-based CS, respectively. We note that these displayed images are the centres of these reconstructed 
images. The model images are displayed with the range 0 Jy/pixel to 0.006 Jy/pixel and the scaling power of $-1.5$; the 
restored images are displayed with the range 0 Jy/pixel to 0.3 Jy/pixel and the scaling power of $-1.5$; the residual images in figure~\ref
{fig_none} are truncated and displayed with range -0.01 Jy/pixel to +0.01 Jy/pixel and the scaling power of 0. From these reconstructed models, we can see that IUWT-based CS gives the best model reconstruction and residual image. Not unexpectedly, there are some point-like structures in the model of the H\"ogbom CLEAN. The multiscale CLEAN gives a smooth version of the model and a non-uniform residual image. The PF is good for this case, because of the large number of measurements. 
When we turn to the restored images, it is difficult to discern significant differences visually. Consequently, numerical comparison is important. The following numerical 
comparisons are carried out with dynamic range (DR) and fidelity. 
As defined in~\citep{Cornwell:1993p1113}, the dynamic range describes the ratio of the peak brightness of the restored image to the off-source error level and is defined as
\begin{equation}
\label{equ_dr}
DR=\frac{max(restored \;image)}{RMS \;error},
\end{equation} where the root mean square (RMS) error is defined to be
\begin{equation}
\label{equ_rms}
RMS \;error=\sqrt {\frac {\sum(residual\; image)^2}{number \;of \;pixels }}.
\end{equation}

The fidelity (FD) is the ratio of the brightness of a pixel in the true image to the error image. The fidelity is an image that is difficult to 
measure. Therefore, the simplified definition is adopted

\begin{equation}\label{equ_fd}
FD= median \{\frac{true \; sky \; image \; image}{abs(model-true \; sky \; image)}\}.
\end{equation}

This definition slightly differs from the one defined in~\citet{Cornwell:1993p1113}. The numerator is the true sky image rather than the model, and Eq. (\ref{equ_fd}) provides a more reliable evaluation than the one defined in~\citet{Cornwell:1993p1113}. If there are two different models with the same denominator, these two different models will have the same FD in Eq. (\ref{equ_fd}). However, these two different models will have different FDs in the study of~\citet{Cornwell:1993p1113}, which is undesirable. Therefore, Eq. (\ref{equ_fd}) is adopted in this paper. To achive a robust evaluation of the restored model, we propose the clean beam blurred FD (CFD) given by 

\begin{table*}%[h]
\caption{Numerical comparison results}
\begin{center}
\begin{tabular}{ccccccc}
%\begin{tabular}{|c|c|c|c|c|c|c|c|c|} %% this creates two columns
% |l|l| to left justify each column entry
% |c|c| to center each column entry
% use of \rule[]{}{} below opens up each row
   \hline
            \noalign{\smallskip}
                   & H\"ogbom     & Multiscale     &PF       & IUWT-based CS\\
   \noalign{\smallskip}
            \hline
            \noalign{\smallskip}
    \textbf{Uniform-weighted UV coverage} & & & &&& \\
   \noalign{\smallskip}
            \hline
            \noalign{\smallskip}
   DR       &    188     & 166  &      154	& 186\\
  FD          &  1.292	 &  2.337 &     1.965 &   2.569\\
  CFD  & 1.001 &	1.014&	1.014&	1.035 \\
  Time (minutes)    &    34    & 17 &  1	& 3\\
     \noalign{\smallskip}
            \hline
            \noalign{\smallskip}
    \textbf{Natural-weighted UV coverage} & & & &&& \\
    \noalign{\smallskip}
            \hline
            \noalign{\smallskip}
  DR       &    396    & 1477  &    834	&  1175\\
  FD          &  1.000	 &  2.812&    8.295 &   8.786\\
  CFD  & 0.976 &	1.308&	3.319&	3.367 \\
  Time (minutes)    &    31   & 18 &  2	& 28 \\    \noalign{\smallskip}
            \hline
\end{tabular}
\end{center}
\label{tab:numbericalcoompare}
\end{table*}

\begin{eqnarray}\label{equ_cfd}
CFD=median \{\frac{clean\;  beam *true \; sky  \; image}{clean \;beam * \;  abs(model-true \; sky \; image)}\}.
\end{eqnarray} This is a blurred version of FD defined in Eq. (\ref{equ_fd}) and will decrease the different performance of the models but more robustly than FD.  
Numerical comparison results can be found in Table~\ref{tab:numbericalcoompare}. From the first part of the table (the results of the uniform weighting test), 
we can see that IUWT-based CS provides the best reconstruction results for both FD and CFD.

\begin{figure*}
 \begin{center}
 \begin{tabular}{cc}
	  \subfigure{
      \includegraphics[scale=0.3]{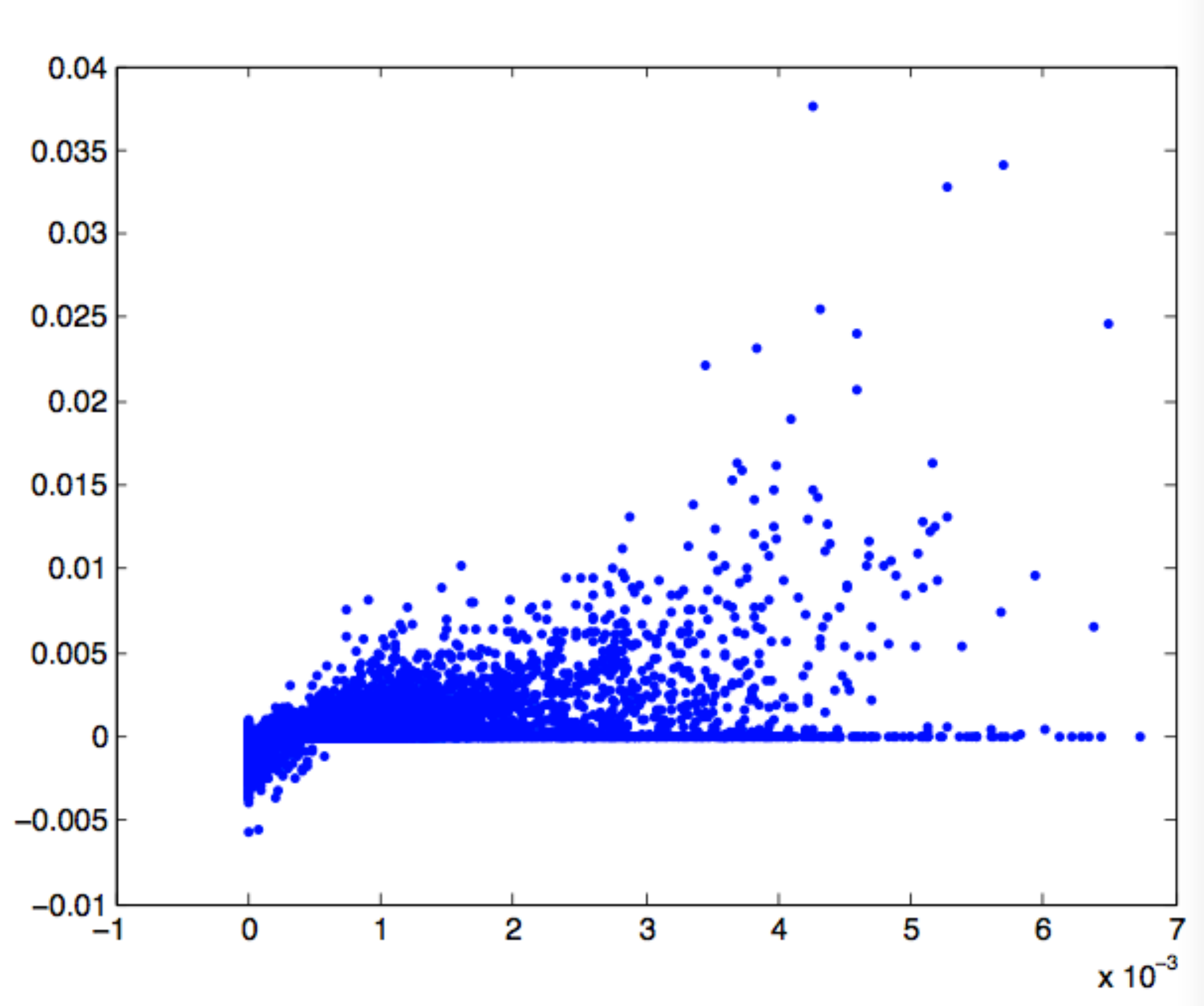}
     \label{fig_nonecleanplot}
     }
      \subfigure {
      \includegraphics[scale=0.3]{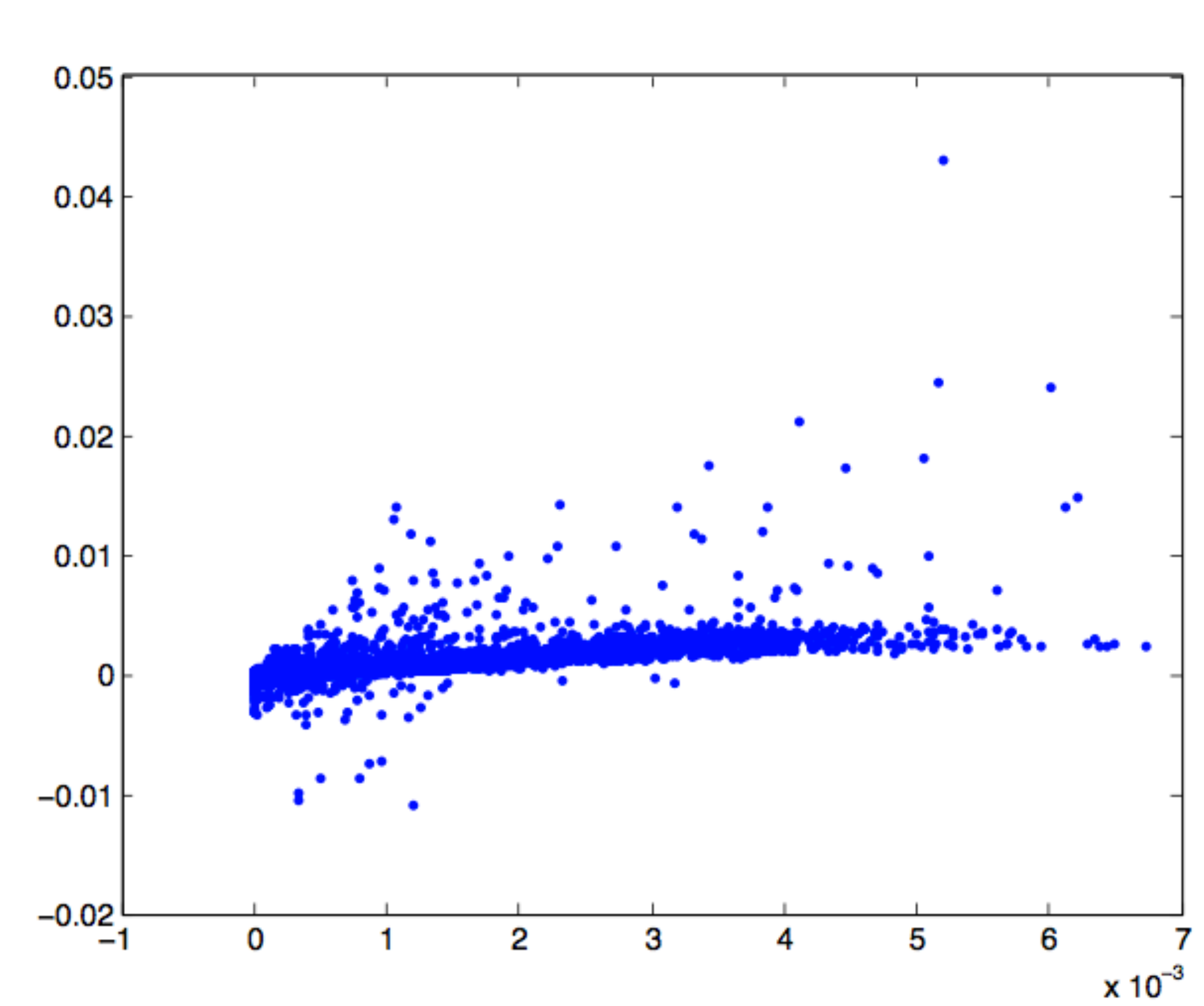}
     \label{fig_nonemscleanplot}
     }
  \end{tabular} 
        \begin{tabular}{cc}
	  \subfigure{
      \includegraphics[scale=0.3]{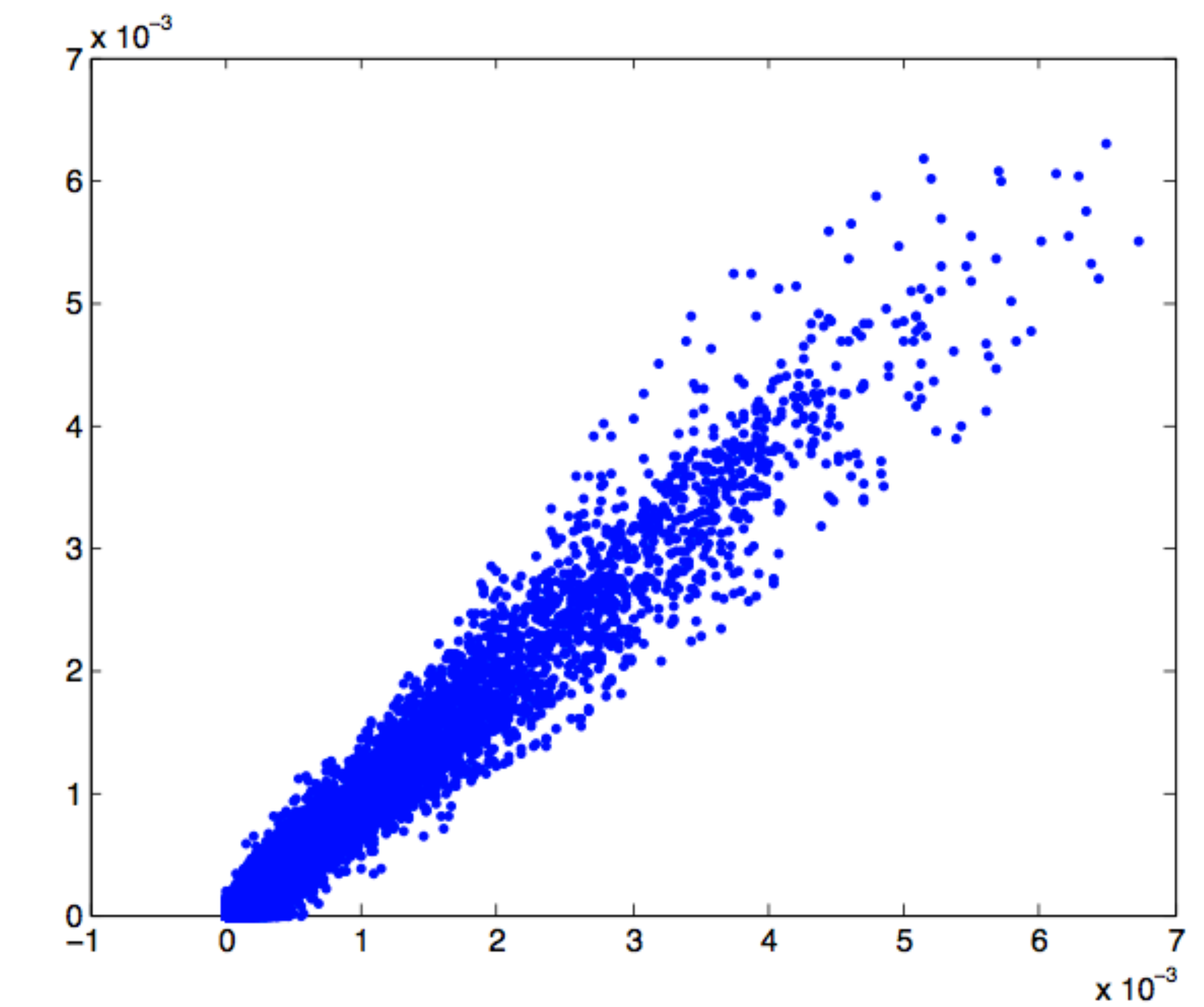}
     \label{fig_nonepfplot}
     }
\hspace{0.08in}
      \subfigure {
      \includegraphics[scale=0.3]{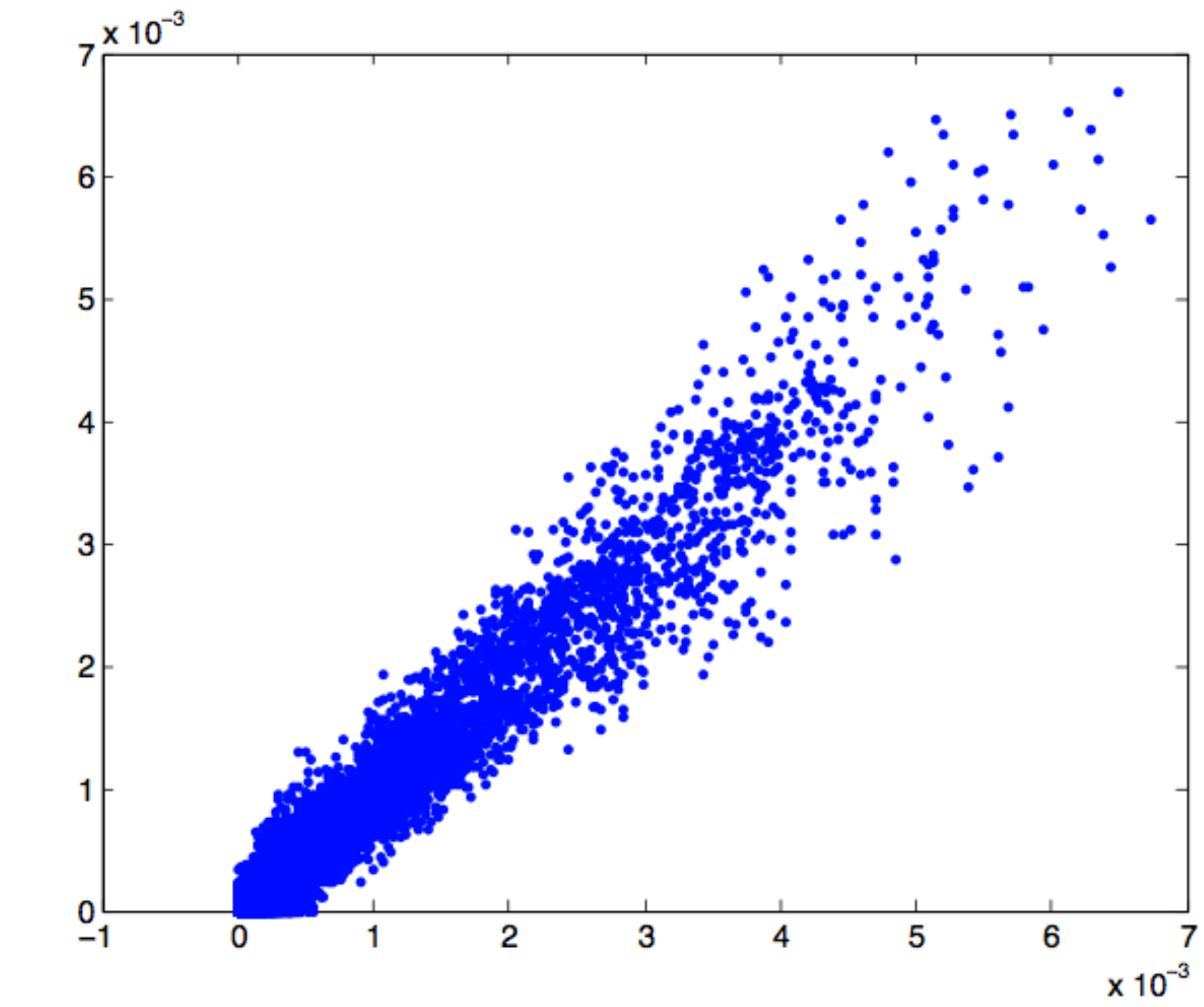}
     \label{fig_noneiuwtplot}
     }
  \end{tabular} 
 \end{center}
\caption{ The plots of the models versus the true sky image with natural weighting:~\subref{fig_nonecleanplot} The H\"ogbom CLEAN, ~\subref{fig_nonemscleanplot} the multiscale CLEAN, ~\subref{fig_nonepfplot}  the partial Fourier (PF) method, and~\subref{fig_noneiuwtplot} the IUWT-based CS. The y-axis is the reconstructed model and the x-axis is the true sky image.
} \label{fig_noneplot}
\end{figure*}

For the natural weighting test, we adopt a 10000 iterations, a gain of 0.1, and a 0.001Jy threshold for the H\"ogbom CLEAN. As far as the 
multiscale CLEAN method is concerned, the following arguments are adopted: 10000 iterations; 
gain of 0.7; six scales as  0, 2, 4, 8, 16, 32, and 0.001Jy threshold. For the CS-based methods, $\lambda$ is set to $0.00001$ for both the PF and the IUWT-based CS.  It takes 17 iterations for the PF to converge and 50 iterations for the the IUWT-based CS. All the deconvolved results are shown in figure~\ref{fig_none}. From left to right in columns are the model images, the 
residual images and the restored images, respectively. From top to bottom in rows are the results of the H\"ogbom CLEAN, the multiscale CLEAN, the partial Fourier (PF) method, and the IUWT-based CS, respectively. Again, these displayed images are the centre of those reconstructed 
images. As in the previous figure, the model images are displayed with range 0 Jy/pixel to 0.006 Jy/pixel and the scaling power of $-1.5$;  and the 
restored images are displayed with the range 0 Jy/pixel to 0.3 Jy/pixel and the scaling power of $-1.5$; the residual images in figure~\ref
{fig_none} are truncated and displayed with the range of from -0.01 Jy/pixel to +0.01 Jy/pixel and the scaling power of 0. 
 
From figures~\ref{fig_none}, we can see that the IUWT-based CS can provide superior 
results to the H\"ogbom CLEAN, the multiscale CLEAN, and the PF. For example, the model of the IUWT-based CS is clearly seen to be a closer approximation to the true sky image when comparing with other methods. The residual image of the IUWT-based CS is also the most uniform one in the middle column of figure~\ref{fig_none}, and the restored image of the IUWT-based CS shows fewer artifacts than those of the other methods.

As well as dynamic range and fidelity, it is important to study the photometry $i.e.$ the relationship between the true sky pixel and that reconstructed. Ideally this should be perfectly linear with slope unity, becoming more scattered only for noisy pixels. We show the photometric curves for the full resolution images in figure~\ref{fig_none}. We plot all the models against the true sky image in figure~\ref{fig_noneplot}. All models are displayed along the y-axis, and the x-axis is the true sky image. From left to right and from top to bottom, we provide plots of the H\"ogbom CLEAN, the multiscale CLEAN, the partial Fourier (PF) method, and the IUWT-based CS. Here, we can see that both the PF and the IUWT-based CS give a much closer approximation to the true sky image. Note that we do not display those plots for the uniform weighting test, because there is almost no difference between those plots for that test. If we plot similar curves for the restored images, we find much smaller differences between the methods, confirming the wisdom of restoring the CLEAN-based images to a lower resolution. If, for scientific reasons, higher resolution is required, then the CS-based approaches should be used.

Numerical comparison results of natural weighting test can also be found in the second part of Table~\ref{tab:numbericalcoompare}. From this table, 
we can see that IUWT-based CS again provides the best reconstruction, even though it has a lower dynamic range than the multiscale CLEAN. 
As far as the computational time is concerned, in the uniform weighting test, the IUWT-based CS is much faster than traditional deconvolution methods. It is slower for the natural weighting test, but it is still comparable in speed to the H\"ogbom CLEAN. Since these CS-based deconvolution methods are gradient-based methods, they will not heavily depend on the complexity of the model. This is the reason why they are faster than the H\"ogbom CLEAN. The computer is a 2.53-GHz Core 2 Duo MacBook Pro with 4GB RAM. 

In both cases, IUWT-based CS can reconstruct good results as judged visually and numerically.  The different weighting methods do not affect the good performance. Our interpretation is that the IUWT-based CS provided the best results because IUWT provides a more sparse representation. If the target signal consists of point sources, then PF will provide superior deconvolution results.

\section{Conclusion}
\label{sec:conclusion} 
%Compressive sensing will bring great opportunity to the design of a radio telescope array and the 
%deconvolution methods for a given radio telescope array. 

For the radio interferometry deconvolution problem, CS-based methods can provide better reconstructions than the traditional 
deconvolution methods for uniform weighting or natural weighting. In general, for 
point sources, PF is the best approach; for extended sources, the IUWT decomposition-based 
deconvolution method (called IUWT-based CS in this paper) is a good solution. 
The precondition is 
we should know some prior knowledge of the target signal, for example, we need to know in 
which domain the target signal has a sparse representation. In some circumstances, this prior 
information might not be available, therefore, future work will focus on an adaptive 
deconvolution method for any sources. The other potential project is to implement these methods for deconvolving those large-scale images in real time. To achieve this goal, this algorithm can be run on clusters by cutting the large-scale images into blocks in order to carry out parallel computing. 

\begin{acknowledgements}
We thank Jean-Luc Starck, Yves Wiaux, and Alex Grant for very helpful discussions concerning
CS methods. In particular, we thank Jean-Luc Starck and Arnaud Woiselle for the inpainting C++ toolbox (private communication).
The PF has been built into the ASKAP software package and the IUWT-based CS is on the way .
\end{acknowledgements}

\bibliographystyle{aa}
 \bibliography{csiro_bibtex.bib}

\end{document}